\documentclass[pdflatex,sn-mathphys-num]{sn-jnl}
\usepackage{inconsolata}
\usepackage{graphicx}
\usepackage{multirow}
\usepackage{amsmath,amssymb,amsfonts}
\usepackage{amsthm}
\usepackage{mathrsfs}
\usepackage[title]{appendix}
\usepackage{xcolor}
\usepackage{subcaption}
\usepackage{textcomp}
\usepackage{manyfoot}
\usepackage{booktabs}
\usepackage{algorithm}
\usepackage{algorithmicx}
\usepackage{algpseudocode}
\usepackage{listings}
\usepackage{braket}
\usepackage{siunitx}
\usepackage{quantikz}

\theoremstyle{thmstyleone}

\theoremstyle{thmstyletwo}%

\theoremstyle{thmstylethree}%

\renewcommand{\theequation}{\thesection.\arabic{equation}}

\begin{document}

\title[]{Full Quantum Stack: Ket Platform} 

\author[1]{\fnm{Evandro} Chagas Ribeiro da \sur{Rosa}}

\author[1]{\fnm{Eduardo} Willwock \sur{Lussi}}

\author[1]{\fnm{Jerusa} \sur{Marchi}}

\author[1]{\fnm{Rafael} \sur{de Santiago}}

\author[2]{\fnm{Eduardo} Inacio \sur{Duzzioni}}

\affil[1]{\orgdiv{Departamento de Informática e Estatística}, \orgname{Universidade Federal de Santa Catarina}, \orgaddress{\city{Florianópolis}, \state{Santa Catarina}, \country{Brazil}}}

\affil[2]{\orgdiv{Departamento de Física}, \orgname{Universidade Federal de Santa Catarina}, \orgaddress{\city{Florianópolis}, \state{Santa Catarina}, \country{Brazil}}}

\abstract{As quantum computing hardware continues to scale, the need for a robust software infrastructure that bridges the gap between high-level algorithm development and low-level physical qubit control becomes increasingly critical. A full-stack approach, analogous to classical computing, is essential for managing complexity, enabling hardware-agnostic programming, and systematically optimizing performance. In this paper, we present a comprehensive, end-to-end quantum software stack, detailing each layer of abstraction from user-facing code to hardware execution. We begin at the highest level with the Ket quantum programming platform, which provides an expressive, Python-based interface for algorithm development. We then describe the crucial multi-stage compilation process, which translates hardware-agnostic programs into hardware-compliant circuits by handling gate decomposition, qubit mapping to respect device connectivity, and native gate translation. To illustrate the complete workflow, we present a concrete example, compiling the Grover diffusion operator for a superconducting quantum processor. Finally, we connect the compiled circuit to its physical realization by explaining how native gates are implemented through calibrated microwave pulses. This includes the calibration of single- and two-qubit gates, frequency characterization, and measurement procedures, providing a clear picture of how abstract quantum programs ultimately map onto the physical control of a quantum processor. By providing a detailed blueprint of a complete quantum stack, this work illuminates the critical interplay between software abstractions and physical hardware, establishing a framework for developing practical and performant quantum applications.}

\keywords{Quantum Software Stack, Quantum Compilation, High-Level Quantum Programming, Quantum Control, Quantum Hardware, Ket Platform}

\maketitle

\tableofcontents

\section{Introduction}\label{sec:introduction}

Quantum computers have the potential to solve problems that are intractable for their classical counterparts~\cite{shorPolynomialTimeAlgorithmsPrime1997}. While current quantum hardware has demonstrated a quantum advantage~\cite{aruteQuantumSupremacyUsing2019,madsenQuantumComputationalAdvantage2022,zhongQuantumComputationalAdvantage2020,wuStrongQuantumComputational2021,zhongPhaseProgrammableGaussianBoson2021,Pokharel2023,Liu2025}, the development of large-scale, practical quantum computers remains an open challenge~\cite{Preskill2012,GoogleQuantumAIandCollaborators2025}. Achieving this objective requires significant improvements in quantum hardware engineering. However, a frequently overlooked aspect is the software infrastructure required to manage this hardware.

Although drag-and-drop interfaces for quantum gates provide an introductory approach to quantum computing, coding is essential for developing practical quantum software, even for today's Noisy Intermediate-Scale Quantum (NISQ) computers~\cite{preskillQuantumComputingNISQ2018}. Many quantum programming platforms~\cite{Javadi-Abhari2024,svoreEnablingScalableQuantum2018,steigerProjectQOpenSource2018,bichselSilqHighlevelQuantum2020,Smith2016,sivarajahT|ketRetargetableCompiler2021,Efthymiou2022}, though sometimes accessible via high-level languages, primarily offer low-level quantum constructs, which are currently necessary to obtain meaningful results from noisy quantum devices. However, as the number of qubits increases, the time required for low-level programming and qubit-specific tuning becomes impractical; also, the resulting code often lacks hardware independence. Consequently, there is a strong motivation to adopt a higher-level programming model, where automated compiler optimizations replace manual, hardware-specific tuning.

The transition from low-level to high-level coding is not novel; a similar evolution occurred in classical computing. Although the C language was invented in 1972~\cite{Kernighan1991} and standardized by the American National Standards Institute (ANSI) in 1989, assembly coding\footnote{The human-readable language closest to machine code.} was often used for performance-critical applications until the late 1990s. With advancements in compilers and the increasing complexity of processors, writing assembly code that could outperform compiler-generated code became increasingly challenging, as compilers could explore a wider range of optimizations than the average programmer. Consequently, high-level programming enhances both programmer productivity and execution performance. Just as classical computing performance has advanced through both hardware and software improvements, a similar trajectory is expected for quantum computing.

In this paper, we review the software infrastructure required to execute a quantum program written in Python on quantum hardware, presenting the components of a complete \emph{quantum stack}. A software stack is a computer science concept that divides a solution into abstraction layers, starting with the problem domain at the highest level and becoming more specialized at lower levels. Our proposed quantum stack begins at a high level with Python code using the Ket quantum programming platform~\cite{darosaKetQuantumProgramming2022} and extends down to the microwave pulses that manipulate the physical qubits.

The Ket platform is an open-source project composed of three main components. The first is a Python programming interface, which serves as the high-level front-end for quantum software development. The second is Libket, the platform's runtime library, which is responsible for quantum compilation; this runtime is written in Rust, exposes a C API, and can be used independently of the Python interface. The third component is the Ket Bitwise Simulator (KBW), a noiseless simulator that enables the execution of quantum applications on ordinary hardware.

This review addresses the stack layers currently implemented in Ket, from the programming interface down to the low-level quantum circuit. The broader objective, however, is to expand this stack in both directions: upwards with domain-specific libraries for fields such as quantum chemistry~\cite{Tilly2022} and finance~\cite{hermanQuantumComputingFinance2023}, and downwards to include pulse-level programming~\cite{alexanderQiskitPulseProgramming2020} and direct Arbitrary Waveform Generator (AWG) management~\cite{stefanazziQICKQuantumInstrumentation2022}.

Structuring the software infrastructure into abstraction layers helps manage complexity, allowing developers to focus on one layer at a time. Furthermore, this layered approach facilitates the integration of new solutions and optimizations. This is because modifications typically require a new component to interact only with the immediately adjacent layers, rather than demanding a rewrite of the entire software stack, an aspect discussed in subsequent sections.

The remainder of this paper is structured as follows. Section~\ref{sec:quantum-programming} provides an overview of the Ket quantum programming platform, which enables the development of quantum software using Python. Within the quantum stack, Ket represents the highest abstraction layer, which is completely agnostic to the target quantum hardware and is used by programmers to develop domain-specific solutions. Section~\ref{sec:compiler} describes how quantum code written in Ket is compiled for a target quantum computer, ensuring the final circuit adheres to the hardware's connectivity constraints and native gate set. The compilation process concludes by scheduling the pulses that physically manipulate the qubits, a step that depends on calibration parameters measured from the quantum device. Section~\ref{sec:pulse} explores the methods and hardware aspects involved in calibrating a quantum computer to enable the execution of quantum software. We conclude in Section~\ref{sec:conclusion} with our final remarks.

\section{Quantum Programming Guide}\label{sec:quantum-programming}

Ket is an open-source quantum programming platform designed to bring the expressivity of Python to quantum programming. Its core, \emph{Libket}, is a high-performance runtime library written in Rust, a memory safety language known for its speed. This two-language architecture provides Python's ease of use for development and Rust's performance for computationally intensive tasks. The platform also includes a built-in quantum simulator, named \emph{KBW}, which allows users to simulate small-scale quantum programs on a personal computer.

This section serves as a programming guide with Ket, demonstrating how to construct quantum programs in Python. It assumes the reader is familiar with both Python and the fundamentals of quantum computing, as our focus is on Ket's specific functions and programming model. We will cover the entire workflow, from setting up the execution environment and allocating qubits to applying quantum gates and extracting classical results through measurement.

For those wishing to execute the code examples, Ket can be installed from the Python Package Index (PyPI). It is compatible with Python 3.10 or newer and is available for Linux, Windows, and macOS (Intel and Apple silicon). Ket can be installed via pip:
\begin{center}
    \texttt{pip install ket-lang==0.9.1}
\end{center}
For users on non-x86\_64 processors (such as ARM) under Linux or Windows, the Rust compiler is required to build Ket from source. The examples in this guide use Ket version 0.9, which was the latest version at the time of writing. For complete documentation, please visit \url{https://quantumket.org}.

\subsection{Classical-Quantum Model}

\begin{figure}[htpb]
    \centering
    \includegraphics[width=0.8\linewidth]{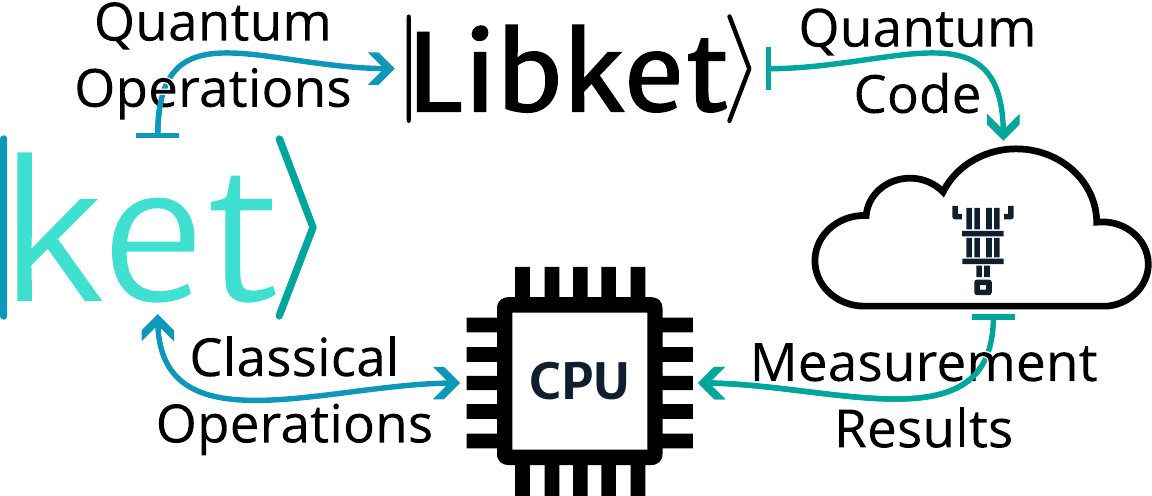}
    \caption{The hybrid computational model in Ket. The programmer defines operations in Python. Classical logic is executed by the CPU, while quantum instructions are collected by the Libket runtime into a \emph{quantum code}. This code is sent to a backend (a simulator or QPU), which returns classical measurement results. These results can then be used to guide the subsequent steps of the algorithm, forming a complete classical-quantum feedback loop.}
    \label{fig:ket:runtime}
\end{figure}

Quantum software solutions are inherently hybrid programs, where the quantum computer acts as a specialized accelerator, much like a Graphics Processing Unit (GPU) or a Field Programmable Gate Array (FPGA). As illustrated in Figure~\ref{fig:ket:runtime}, when a quantum program is written in Ket, a clear division of labor occurs: classical instructions are treated as standard Python code executed primarily by the CPU, while quantum operations are passed to Libket, which is responsible for managing the quantum execution.

Libket gathers these operations into a \emph{quantum code}--an internal representation of a quantum circuit--and dispatches it to the target backend, which can be a simulator or a quantum computer. Once the execution is complete, the backend returns classical measurement results to the Python program. This allows the classical code to process the quantum output and determine the final answer.

We highlight the classical computer's role in interpreting the output of a quantum computation. A quantum device typically produces raw classical data, which must be post-processed to extract a final solution. Shor's algorithm serves as a prime example. After the quantum processor generates a bitstring related to a problem's period, the classical computer computes the continued fraction algorithm to determine the period.

To manage the classical-quantum interaction, Ket provides two execution modes that differ in how the two processors interact. In \emph{batch execution} mode, all quantum operations are collected by Libket into a complete program before being sent to the target. This program is dispatched only when a result is requested. On the backend, batch execution allows requests to be managed by a queue, which reduces the quantum computer's idle time and optimizes resource usage. This model is also more efficient for NISQ devices as it is not impacted by classical-quantum communication latency.

In \emph{live execution} mode, each quantum instruction is sent immediately to the execution target. Measurement results are available instantly, allowing classical logic to influence subsequent quantum operations. This dynamic control is essential for implementing protocols like quantum teleportation and error correction schemes~\cite{devittQuantumErrorCorrection2013,Chatterjee2023}. However, this interaction introduces a latency that is often impractical for current NISQ hardware. Consequently, this mode is primarily used with local simulators and is helpful for the study and debugging of quantum algorithms.

\subsection{Quantum Process and Qubit Allocation}\label{subsec:ket:process}

The entry point for any quantum program in Ket is the \texttt{Process} class, which manages the quantum execution context. It encapsulates all necessary information for execution, including the number of available qubits, the native gate set of the target device, qubit connectivity, and measurement capabilities.

Once a \texttt{Process} instance is created, its primary use is to allocate qubits via the \texttt{.alloc()} method. This method takes an optional integer to specify the number of qubits to allocate; if omitted, it allocates a single qubit. The method returns a \emph{Quant} object, which is an opaque data type that acts as a list of qubit references. Newly allocated qubits are always initialized in the $\ket{0}$ state.

The default \texttt{Process()} constructor initializes the KBW simulator with 32 qubits available in sparse mode (see Section~\ref{subsubsec:kbw_config} for details). This provides a convenient starting point for developing and testing quantum programs. Figure~\ref{fig:ghz_preparation} shows how to instantiate a \texttt{Process} and allocate 30 qubits to prepare a GHZ state. Multiple calls to \texttt{.alloc()} are allowed as long as qubits remain available.

\begin{figure}[htbp]
    \centering
    \begin{minipage}{.5\linewidth}
        \begin{lstlisting}[language=Python]
process = Process() 
qubits = process.alloc(30)
H(qubits[0]) 
ctrl(qubits[0], X)(qubits[1:])
dump(qubits).show()
        \end{lstlisting}
    \end{minipage}
    \caption{Example of qubit allocation using a \texttt{Process} instance and the subsequent preparation of a 30-qubit GHZ state.}
    \label{fig:ghz_preparation}
\end{figure}

The \texttt{Quant} object holds references to \emph{logical qubits}, which are indexed from 0 to $N-1$ where $N$ is the total number of allocated qubits. These logical indices do not necessarily correspond to the physical qubit labels on a Quantum Processing Unit (QPU). The compiler may map a single logical qubit to different physical qubits during execution to satisfy the hardware's connectivity constraints.

\subsection{Quantum Gates}\label{subsec:ket:gates}

In Ket, quantum gates are represented as Python functions that take qubit references as arguments. Ket is flexible: any Python callable can act as a quantum gate, so long as it does not measure or allocate qubits\footnote{This excludes the concept of ancillary qubits, which is not addressed in this paper.}. This design simplifies the creation of new quantum operations and allows for the straightforward application of inverse and controlled modifiers to any gate-like function.

\begin{figure}[htbp]
    \centering
    \begin{subfigure}[b]{.49\linewidth}
        \begin{lstlisting}[language=Python]
def example(qubits: Quant):
    H(qubits[0])
    RY(pi / 3, qubits[:2])
    X(qubits)

# Args: function, num_qubits
qulib.draw(example, 3)
        \end{lstlisting}
        \caption{Python code defining a quantum circuit.}
        \label{fig:circuit_visualization_code}
    \end{subfigure}
    \hfill
    \begin{subfigure}[b]{.49\linewidth}
        \centering
        \includegraphics[scale=.65]{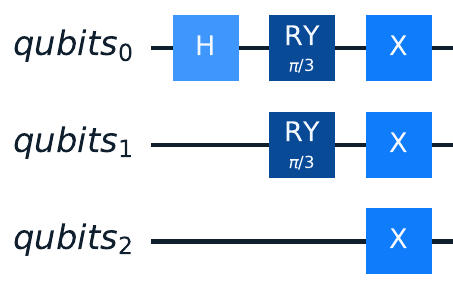}
        \caption{Resulting quantum circuit diagram.}
        \label{fig:circuit_visualization_diagram}
    \end{subfigure}
    \caption{Visualization of a quantum circuit defined in Ket using \texttt{qulib.draw}.}\label{fig:circuit_visualization}
\end{figure}

All quantum gates are ultimately constructed from a set of fundamental single-qubit gates provided by Libket~\cite{rosaOptimizingGateDecomposition2025}, as detailed in Table~\ref{tab:base-gates}. Multi-qubit gates are typically built by applying these fundamental gates in a controlled manner. To visualize a circuit without executing it, Ket provides the \texttt{qulib.draw} function, as shown in Figure~\ref{fig:circuit_visualization}.

Qubits in Ket are managed through the \texttt{Quant} data type. As a list-like object, \texttt{Quant} supports indexing, slicing, and concatenation, with the result of these operations also being a \texttt{Quant} instance. As demonstrated in Figure~\ref{fig:circuit_visualization_code}, applying a single-qubit gate to a multi-qubit \texttt{Quant} object applies the gate to each qubit individually.

\begin{table}[htbp]
    \centering
    \caption{Single-qubit gates provided by Libket.}
    \label{tab:base-gates}
    \begin{tabular}{llc}
        \toprule
        Quantum Gate & Python Function           & Gate Matrix                                                                                                                            \\
        \midrule
        Pauli-X      & \texttt{X(qubit)}         & $\begin{bmatrix} 0 & 1 \\ 1 & 0 \end{bmatrix}$                                                                                         \\ \addlinespace
        Pauli-Y      & \texttt{Y(qubit)}         & $\begin{bmatrix} 0 & -i \\ i & 0 \end{bmatrix}$                                                                                        \\ \addlinespace
        Pauli-Z      & \texttt{Z(qubit)}         & $\begin{bmatrix} 1 & 0 \\ 0 & -1 \end{bmatrix}$                                                                                        \\ \addlinespace
        Hadamard     & \texttt{H(qubit)}         & $\frac{1}{\sqrt{2}} \begin{bmatrix} 1 & 1 \\ 1 & -1 \end{bmatrix}$                                                                     \\ \addlinespace
        X-Rotation   & \texttt{RX(angle, qubit)} & $\begin{bmatrix} \cos(\frac{\theta}{2}) & -i\sin(\frac{\theta}{2}) \\ -i\sin(\frac{\theta}{2}) & \cos(\frac{\theta}{2}) \end{bmatrix}$ \\ \addlinespace
        Y-Rotation   & \texttt{RY(angle, qubit)} & $\begin{bmatrix} \cos(\frac{\theta}{2}) & -\sin(\frac{\theta}{2}) \\ \sin(\frac{\theta}{2}) & \cos(\frac{\theta}{2}) \end{bmatrix}$    \\ \addlinespace
        Z-Rotation   & \texttt{RZ(angle, qubit)} & $\begin{bmatrix} e^{-i\theta/2} & 0 \\ 0 & e^{i\theta/2} \end{bmatrix}$                                                                \\ \addlinespace
        Phase        & \texttt{P(angle, qubit)}  & $\begin{bmatrix} 1 & 0 \\ 0 & e^{i\theta} \end{bmatrix}$                                                                               \\
        \bottomrule
    \end{tabular}
\end{table}

\paragraph*{Sequential and Composed Gates}

\begin{figure}[htbp]
    \centering
    \begin{subfigure}[b]{.52\linewidth}
        \begin{lstlisting}[language=Python]
# Method 1: Sequential application
def bell_seq(a: Quant, b: Quant):
    # More explicit qubit args
    CNOT(H(a), b)
# Method 2: Composition
HI = kron(H, I)
BELL = cat(HI, CNOT)
        \end{lstlisting}
        \caption{Code for a Bell state preparation.}
        \label{fig:bell_state_prep_code}
    \end{subfigure}
    \hfill
    \begin{subfigure}[b]{.47\linewidth}
        \centering
        \includegraphics[scale=.7]{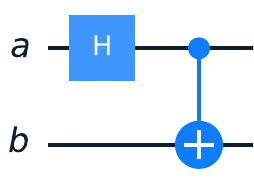}
        \caption{Circuit for a Bell state preparation.}
        \label{fig:bell_state_prep_diagram}
    \end{subfigure}
    \caption{Implementation and visualization of a Bell state preparation using sequential application and gate composition.}\label{fig:bell_state_prep}
\end{figure}

Quantum gates are applied in the order they appear in the Python code. Ket also provides tools to compose gates into more complex operations. Figure~\ref{fig:bell_state_prep} shows two ways to prepare a Bell state. In Ket, gate functions return the qubits they act upon, enabling a fluent interface where gates can be chained in a single line. Alternatively, the \texttt{kron} (tensor product) and \texttt{cat} (concatenation) functions can be used to build composite gates. Gate concatenation with \texttt{cat} is analogous to matrix multiplication, but the gates are listed in the order of their application in the circuit.

\paragraph*{Controlled Operations}

\begin{figure}[htbp]
    \centering
    \begin{subfigure}[b]{.58\linewidth}
        \begin{lstlisting}[language=Python]
def example_cgate(c: Quant, t: Quant):
    # Controlled single gate
    ctrl(c, RX(pi / 2))(t)

    # Controlled user-defined gate
    ctrl(c, bell_seq)(t[0], t[1])
        \end{lstlisting}
        \caption{Defining controlled operations.}
        \label{fig:controlled_gate_example_code}
    \end{subfigure}
    \hfill
    \begin{subfigure}[b]{.41\linewidth}
        \centering
        \includegraphics[scale=.45]{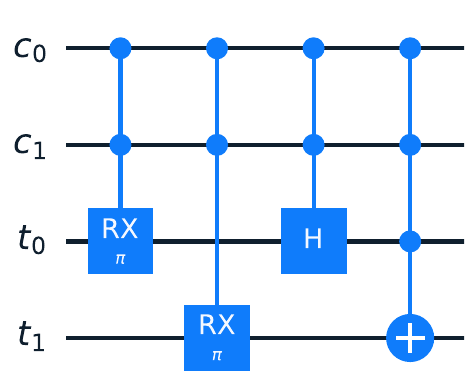}
        \caption{Circuit of controlled operations.}
        \label{fig:controlled_gate_example_diagram}
    \end{subfigure}
    \caption{Constructing controlled gates in Ket using the \texttt{ctrl} function.}\label{fig:controlled_gate_example}
\end{figure}

The CNOT gate is built using Ket's \texttt{ctrl} function, a higher-order function that adds control qubits to any existing gate, including user-defined ones. As shown in Figure~\ref{fig:controlled_gate_example}, when a user-defined function is controlled, the control logic is distributed to each underlying base gate within it.

\paragraph*{Inverse Operations}

\begin{figure}[htbp]
    \begin{subfigure}[b]{.49\linewidth}
        \begin{lstlisting}[language=Python]
def qft(qubits: Quant):
    if len(qubits) == 1:
        H(qubits[0])
        return
    head, *tail = qubits
    H(head)
    for i, c in enumerate(tail):
        angle = 2*pi/2**(i + 2)
        ctrl(c, P(angle))(head) 
    qft(tail)
iqft = adj(qft)
        \end{lstlisting}
        \caption{QFT implementation and its inverse.}
        \label{fig:qft_example_code}
    \end{subfigure}
    \hfill
    \begin{minipage}[b]{.49\linewidth}
        \begin{subfigure}{\linewidth}
            \centering
            \includegraphics[width=\linewidth]{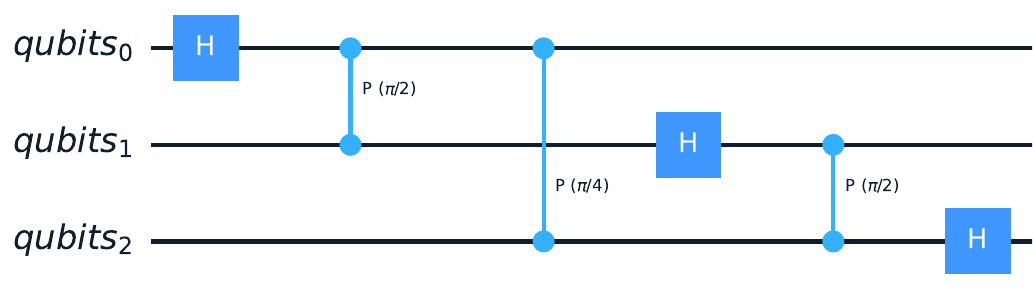}
            \subcaption{QFT circuit diagram.}
            \label{fig:qft_example_diagram_qft}

            \includegraphics[width=\linewidth]{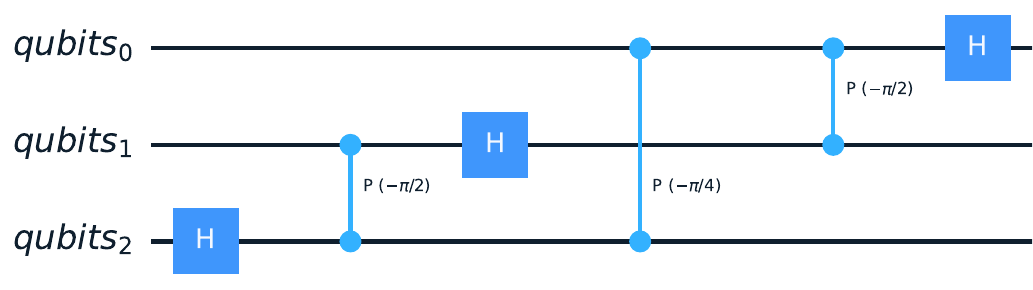}
            \subcaption{Inverse QFT (IQFT) circuit diagram.}
            \label{fig:qft_example_diagram_iqft}
        \end{subfigure}
    \end{minipage}
    \caption{Implementation of QFT and its inverse using the \texttt{adj} function in Ket.}\label{fig:qft_example}
\end{figure}

Every unitary gate $U$ has an inverse, its adjoint $U^\dagger$. Ket provides the \texttt{adj} function to compute the inverse of any gate automatically. Figure~\ref{fig:qft_example} shows a recursive implementation of the Quantum Fourier Transform (QFT). Its inverse (\texttt{iqft}) is simply created by applying \texttt{adj} to the original \texttt{qft} function. The \texttt{adj} function reverses the sequence of operations and takes the adjoint of each individual gate.

\paragraph*{Conjugation by a Unitary}

A common pattern in quantum algorithms is the application of a gate $V$ conjugated by a unitary $U$, forming the structure $U^\dagger V U$. This is analogous to a change of basis. Ket provides the \texttt{with around} construct for this pattern, where the programmer defines $U$ and $V$, and the inverse $U^\dagger$ is applied automatically.

\begin{figure}[htbp]
    \centering
    \begin{subfigure}[b]{.48\linewidth}
        \begin{lstlisting}[language=Python]
# Explicit implementation
def rxx_expl(angle, q0 , q1):
    H(q0) 
    H(q1) 
    CNOT(q0, q1)
    RZ(angle, q1)
    CNOT(q0, q1)
    H(q0) 
    H(q1) 
        \end{lstlisting}
        \caption{Explicit $R_{XX}$ gate implementation.}
        \label{fig:rxx_gate_construction_explicit_code}
    \end{subfigure}
    \hfill
    \begin{subfigure}[b]{.48\linewidth}
        \begin{lstlisting}[language=Python]
# Using 'with around'
def rxx_around(angle, q0, q1):
    H_both = kron(H, H) 
    U = cat(H_both, CNOT)
    with around(U, q0, q1): # U
        RZ(angle, q1)       # V



        \end{lstlisting}
        \caption{$R_{XX}$ gate using \texttt{with around}.}
        \label{fig:rxx_gate_construction_around_code}
    \end{subfigure}
    \begin{subfigure}[c]{\linewidth}
        \centering
        \includegraphics[scale=.7]{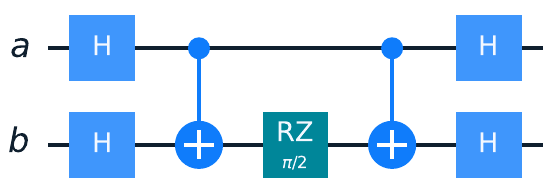}
        \caption{Resulting $R_{XX}$ circuit diagram (identical for both methods).}
        \label{fig:rxx_gate_construction_diagram}
    \end{subfigure}
    \caption{$R_{XX}$ gate implemented explicitly and using Ket's \texttt{with around} construct.}\label{fig:rxx_gate_construction}
\end{figure}

\begin{figure}[htbp]
    \centering
    \begin{subfigure}[b]{.48\linewidth}
        \centering
        \includegraphics[width=\linewidth]{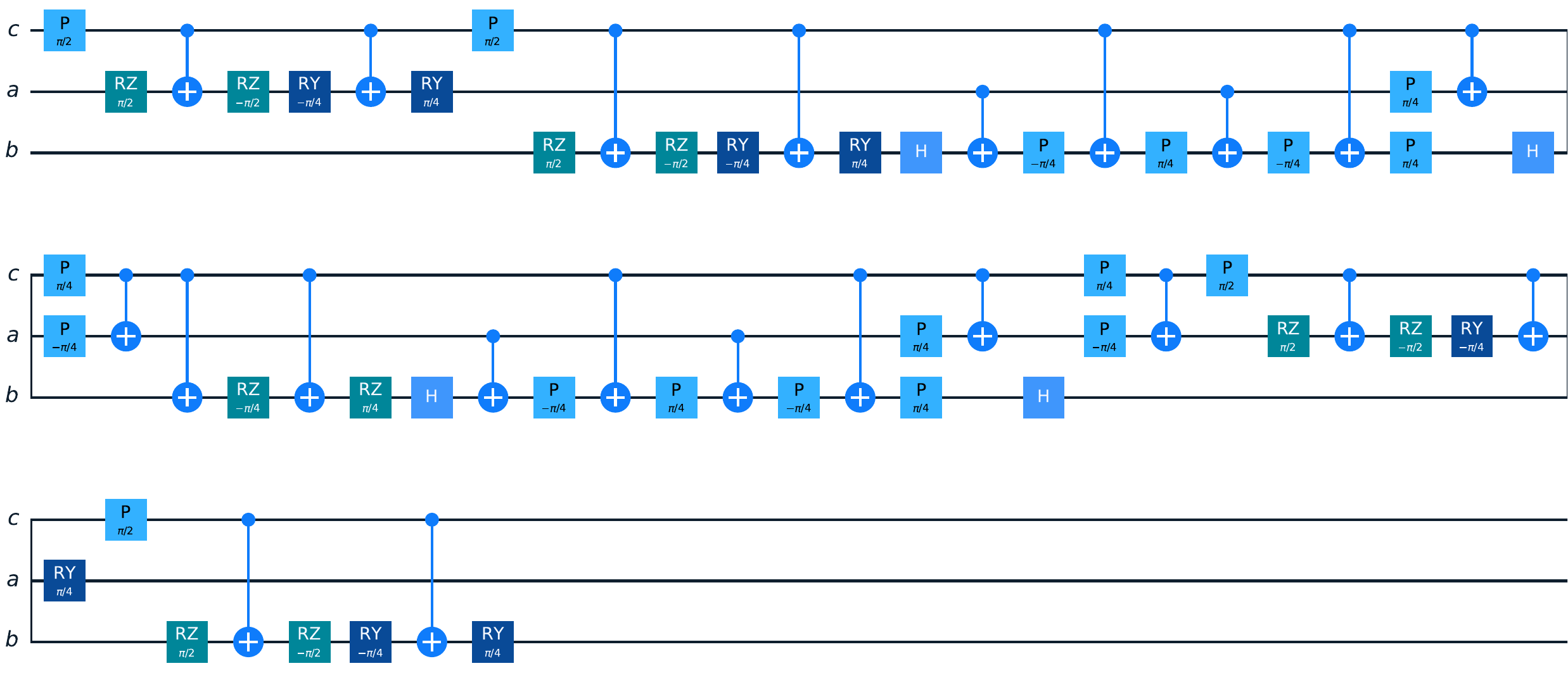}
        \subcaption{Controlled $R_{XX}$ from Figure~\ref{fig:rxx_gate_construction_explicit_code}.}
        \label{fig:controlled_rxx_explicit_decomposed_diagram}
    \end{subfigure}
    \begin{subfigure}[b]{.48\linewidth}
        \centering
        \includegraphics[width=\linewidth]{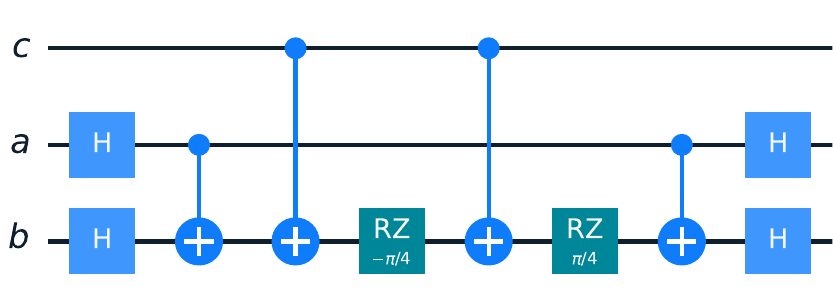}
        \subcaption{Controlled $R_{XX}$ from Figure~\ref{fig:rxx_gate_construction_around_code}.}
        \label{fig:controlled_rxx_around_decomposed_diagram}
    \end{subfigure}
    \caption{Comparison of controlled $R_{XX}$ implementations. The \texttt{with around} construct can lead to more optimized controlled operations, especially upon decomposition.}\label{fig:controlled_rxx_comparison}
\end{figure}

Figure~\ref{fig:rxx_gate_construction} shows two ways to implement an $R_{XX}$ gate. The \texttt{with around} version is more concise and easier to maintain, as changes to $U$ are automatically reflected in both its forward and inverse applications. This construct can also enable compiler optimizations, especially for controlled operations~\cite{rosaOptimizingGateDecomposition2025}. As shown in Figure~\ref{fig:controlled_rxx_comparison}, applying a control to the \texttt{with around} version of $R_{XX}$ can result in a much more efficient circuit upon decomposition compared to controlling the explicit version, reducing the complexity and gate count required for hardware execution~\cite{Rosa2025}.

\subsection{Extracting Information via Measurement}\label{subsec:ket:measure}

Measuring qubits is the fundamental method for extracting classical information from a quantum computation. Ket provides several functions for this purpose, whose availability can depend on the execution target.

\paragraph*{Single-Shot Measurement}
The \texttt{measure(qubits)} function performs a measurement in the computational basis, collapsing the quantum state, as illustrated in Figure~\ref{fig:bell_measure_code}. The result is an object whose \texttt{.get()} method returns an unsigned integer representing the measured bit string. The ability to measure the same qubit multiple times depends on the backend; it is allowed in the KBW simulator but may not be on hardware.

\begin{figure}[htbp]
    \centering
    \begin{minipage}{.7\linewidth}
        \begin{lstlisting}[language=Python]
def bell_measure(a: Quant, b: Quant) -> int:
    CNOT(H(a), b)
    # Measures a and b together
    m = measure(a + b)
    return m.get()
    \end{lstlisting}
    \end{minipage}
    \caption{Example of \texttt{measure}. Output is 0 ($\ket{00}$) or 3 ($\ket{11}$), each with approximately 50\% probability.}
    \label{fig:bell_measure_code}
\end{figure}

\begin{figure}[htbp]
    \centering
    \begin{subfigure}[b]{.55\linewidth}
        \begin{lstlisting}[language=Python]
def bell_sample(a: Quant, b: Quant):
    CNOT(H(a), b)
    s = sample(a + b, 1024)

    # Returns 'Samples' object
    return s
        \end{lstlisting}
        \caption{\texttt{s.get()} example: \texttt{\{0: 529, 3:~495\}}.}
        \label{fig:bell_sample_code}
    \end{subfigure}
    \hfill
    \begin{subfigure}[b]{.44\linewidth}
        \centering
        \includegraphics[scale=.25]{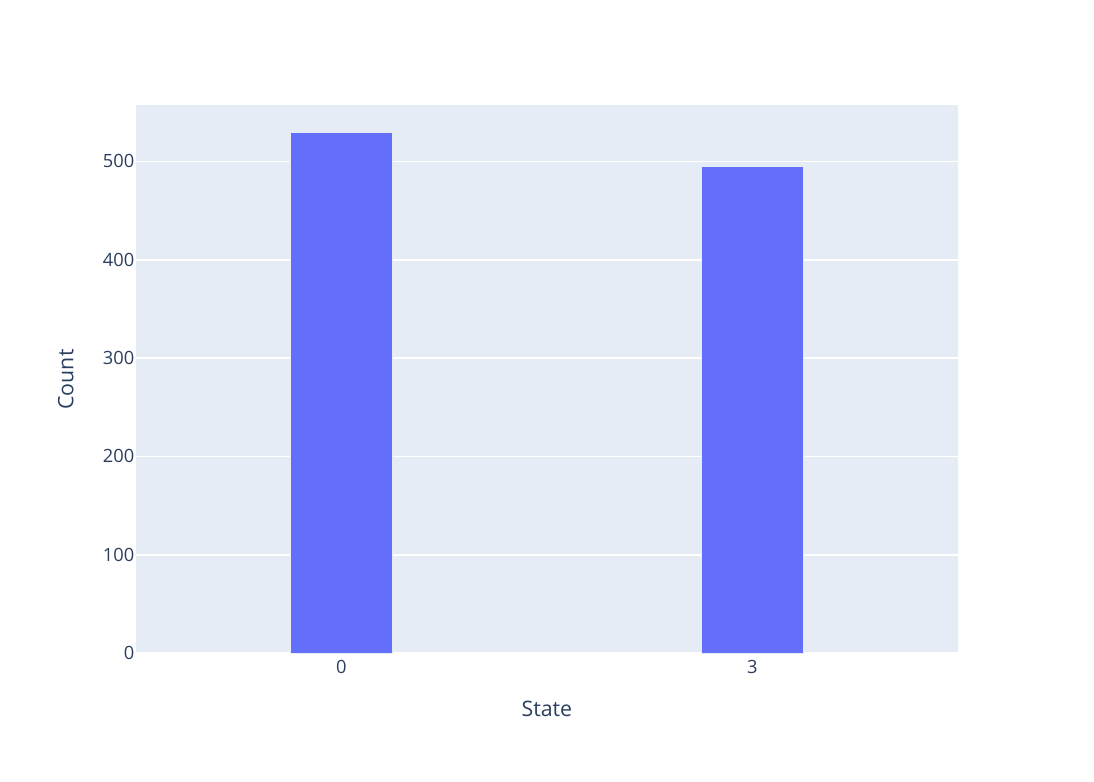}
        \caption{\texttt{s.histogram()} output.}
        \label{fig:bell_sample_hist}
    \end{subfigure}
    \caption{Example of statistical sampling using the \texttt{sample} function and its histogram output. Example with 1024 shots.}
    \label{fig:ket_sample}
\end{figure}

\paragraph*{Statistical Sampling}
The \texttt{sample(qubits, shots)} function measures the qubits \texttt{shots} times and count the results. Its \texttt{.get()} method returns a dictionary mapping measured states to their counts, while \texttt{.histogram()} plots the results, as shown in Figure~\ref{fig:ket_sample}. In the KBW simulator, sampling does not alter the underlying state vector, allowing subsequent operations on the same state. On hardware, sampling consumes the state, requiring re-preparation for further operations.

\paragraph*{Expectation Value Calculation}
The \texttt{exp\_value(o)} function calculates the expectation value of an observable, which is defined using the \texttt{with obs()} context manager by summing weighted products of Pauli operators. Figure~\ref{fig:bell_expv_code} illustrates observable construction and expectation value calculation. The result is a real number retrieved via \texttt{.get()}. Ket also supports automatic gradient calculation for variational algorithms~\cite{Tilly2022}.

\begin{figure}[htbp]
    \centering
    \begin{minipage}{.63\linewidth}
        \begin{lstlisting}[language=Python]
def bell_expv(a: Quant, b: Quant) -> float:
    CNOT(H(a), b)
    with obs():
        A0 = Z(a)
        A1 = X(a)
        B0 = -(X(b)+Z(b))/sqrt(2)
        B1 = (X(b)-Z(b))/sqrt(2)
    # Hamiltonian for CHSH
    h = A0*B0 + A0*B1 + A1*B0 - A1*B1
    ev = exp_value(h)
    return ev.get()
    \end{lstlisting}
    \end{minipage}
    \caption{Example of \texttt{exp\_value} calculating $\braket{h}$ for the CHSH inequality. The expected result is $-2\sqrt{2} \approx -2.8284$.}
    \label{fig:bell_expv_code}
\end{figure}

\begin{figure}[htbp]
    \centering
    \begin{subfigure}[b]{.54\linewidth}
        \begin{lstlisting}[language=Python]
def bell_dump(a: Quant, b: Quant):
    CNOT(H(X(a)), b)
    d = dump(a + b)
    
    # Returns 'QuantumState' object
    return d
        \end{lstlisting}
        \caption{\texttt{d.get()} outputs \texttt{\{0:0.707, 3:-0.707\}}. \texttt{d.show()} displays $\frac{1}{\sqrt{2}}\ket{00} - \frac{1}{\sqrt{2}}\ket{11}$.}
        \label{fig:bell_dump_code}
    \end{subfigure}
    \hfill
    \begin{subfigure}[b]{.45\linewidth}
        \centering
        \includegraphics[scale=.25]{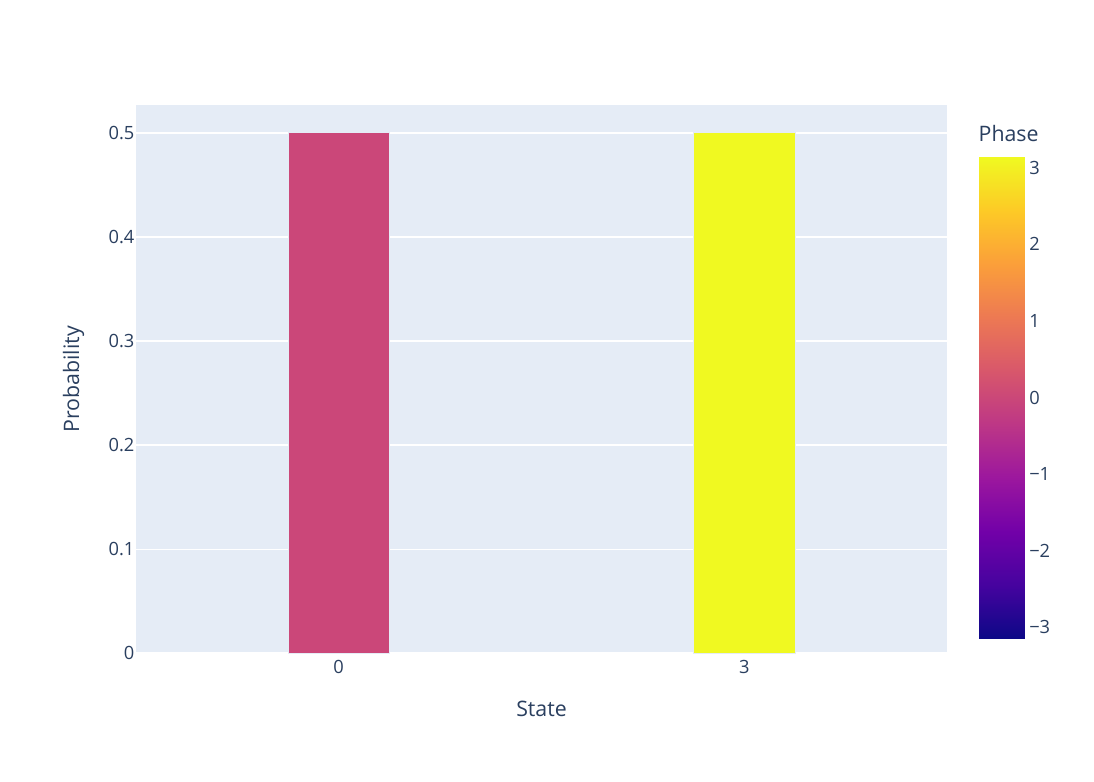}
        \caption{\texttt{d.histogram()} output.}
        \label{fig:bell_dump_hist}
    \end{subfigure}
    \caption{Example of state vector access in a simulator using the \texttt{dump} function.}
    \label{fig:ket_dump}
\end{figure}

\paragraph*{Simulator State Vector Access}
For simulations only, \texttt{dump(qubits)} provides access to the quantum state vector. This is a powerful debugging tool but is not physically possible on a real quantum computer. The returned object's \texttt{.show()} method displays the state in Dirac notation, and \texttt{.get()} returns a dictionary of basis states and their complex amplitudes. Figure~\ref{fig:ket_dump} illustrates this function usage.

\subsection{KBW Simulator Configuration}\label{subsubsec:kbw_config}
The built-in KBW simulator is the default backend and can be configured via keyword arguments in the \texttt{Process} constructor. It has two simulation modes:

\begin{itemize}
    \item The \emph{Dense simulator} is a state-vector simulator that stores the state of $n$ qubits as a complex vector of size $2^n$. It is multithreaded and ideal for algorithms that create highly entangled states, but its memory usage grows exponentially.
    \item The \emph{Sparse simulator} uses a hash map to store only the basis states with non-zero amplitudes. Its performance depends on the number of non-zero amplitudes, making it highly efficient for algorithms where the state remains sparse. For example, it can simulate the 30-qubit GHZ state (Figure~\ref{fig:ghz_preparation}), which has only two non-zero amplitudes, on a standard laptop.
\end{itemize}
The main configuration arguments for the \texttt{Process} constructor are:
\begin{itemize}
    \item \texttt{num\_qubits}: (Integer) Number of qubits. Defaults to 32 for sparse and 12 for dense.
    \item \texttt{simulator}: (String) The simulation mode: \texttt{"sparse"} (default) or \texttt{"dense"}.
    \item \texttt{execution}: (String) The execution mode: \texttt{"live"} (default) or \texttt{"batch"}.
\end{itemize}

\section{Quantum Compilation}\label{sec:compiler}

Analogous to classical high-level programming languages, whose code cannot be directly executed by a CPU, Ket produces high-level quantum circuits that must go through a compilation process to run on a Quantum Processing Unit (QPU). For instance, a program written in Ket may result in circuits with quantum gates that are not natively supported by the QPU or are not in the set of calibrated gates. Furthermore, multi-qubit gates might be applied to qubits that are not physically connected on the device. The compilation process transforms the high-level quantum circuit into one that contains only QPU-supported gates and respects the device's qubit connectivity. This hardware-compliant circuit is then used for pulse scheduling, which generates the sequence of control pulses that physically manipulate the qubits.

\begin{figure}[htbp]
    \centering
    \includegraphics[width=.8\linewidth]{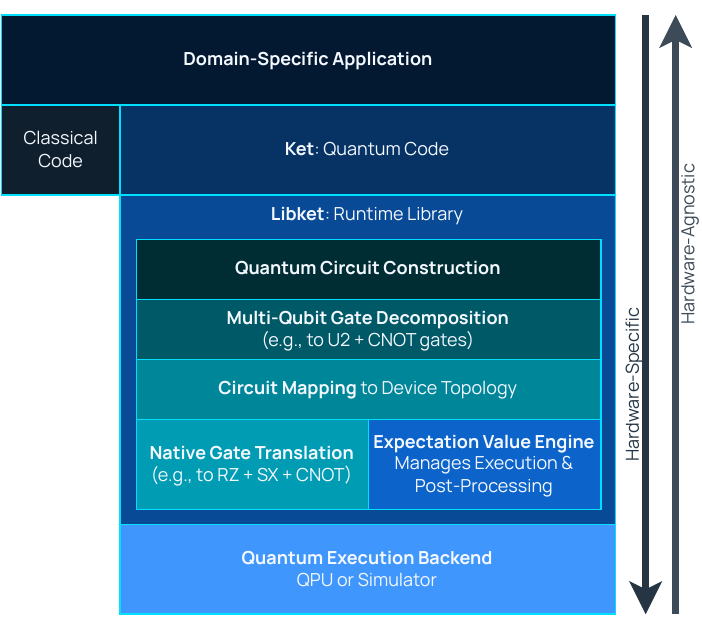}
    \caption{Quantum software stack.}\label{fig:stack}
\end{figure}

Figure~\ref{fig:stack} illustrates the quantum software stack. At the top of the stack is the domain-specific application, which utilizes Ket for quantum development and acceleration. The high-level quantum circuit generated by Ket is then compiled by the platform's runtime library, Libket. This compilation process has three primary stages: multi-qubit gate decomposition (Section~\ref{subsec:decompose}), where multi-controlled gates are broken down into single- and two-qubit gates; quantum circuit mapping (Section~\ref{subsec:mapping}), where logical qubits are mapped to physical qubits respecting the device's connectivity; and native gate decomposition (Section~\ref{subsub:native-gate}), where the single- and two-qubit gates are further translated into the native gate set supported by the QPU.

During compilation, additional processing is required if the desired computational result is an expectation value (Section~\ref{subsub:exp-value}), as different quantum circuit executions are often needed to compute the expected value of a Hamiltonian. Additionally, transverse processes like error mitigation and circuit optimization can be applied throughout the stack to improve the final result.

In this paper, we focus on the quantum software stack for Noisy Intermediate-Scale Quantum (NISQ) computers; therefore, quantum error correction is not treated as a core part of the infrastructure. However, we argue that for future fault-tolerant quantum computers, the higher levels of the stack would remain largely the same. Many of the compilation steps could be repurposed for this new paradigm, with the major change being the addition of a quantum error correction encoding step right before sending the circuit to the quantum computer. Additional processing would also be needed on the quantum computer side for syndrome measurements and the general maintenance of the quantum error correction code.

\begin{figure}[htbp]
    \begin{minipage}[b]{.54\linewidth}
        \begin{subfigure}[b]{\linewidth}
            \centering
            \begin{minipage}{\linewidth}
                \begin{lstlisting}[language=Python]
def diffusion(qubits: Quant):
    with around(cat(H, X), qubits):
        CZ(*qubits)
    \end{lstlisting}
            \end{minipage}
            \caption{Ket code for the Grover diffusion operator.}
            \label{fig:ket:diffusion}
        \end{subfigure}
        \begin{subfigure}[b]{\linewidth}
            \centering
            \includegraphics[width=\linewidth]{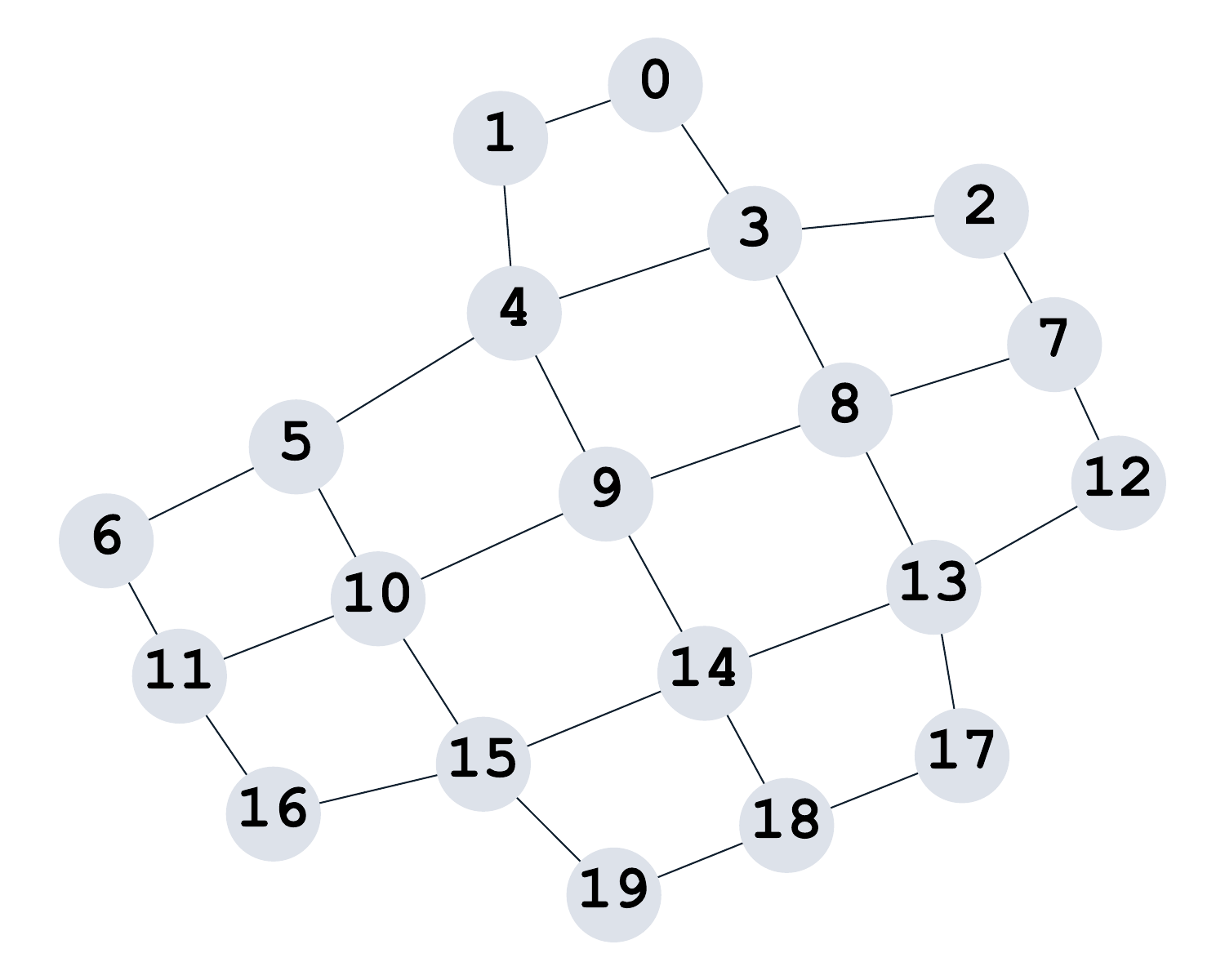}
            \caption{Connectivity of the target QPU.}
            \label{fig:qpu:ex}
        \end{subfigure}
    \end{minipage}
    \hfill
    \begin{subfigure}[b]{.45\linewidth}
        \centering
        \includegraphics[width=.91\linewidth]{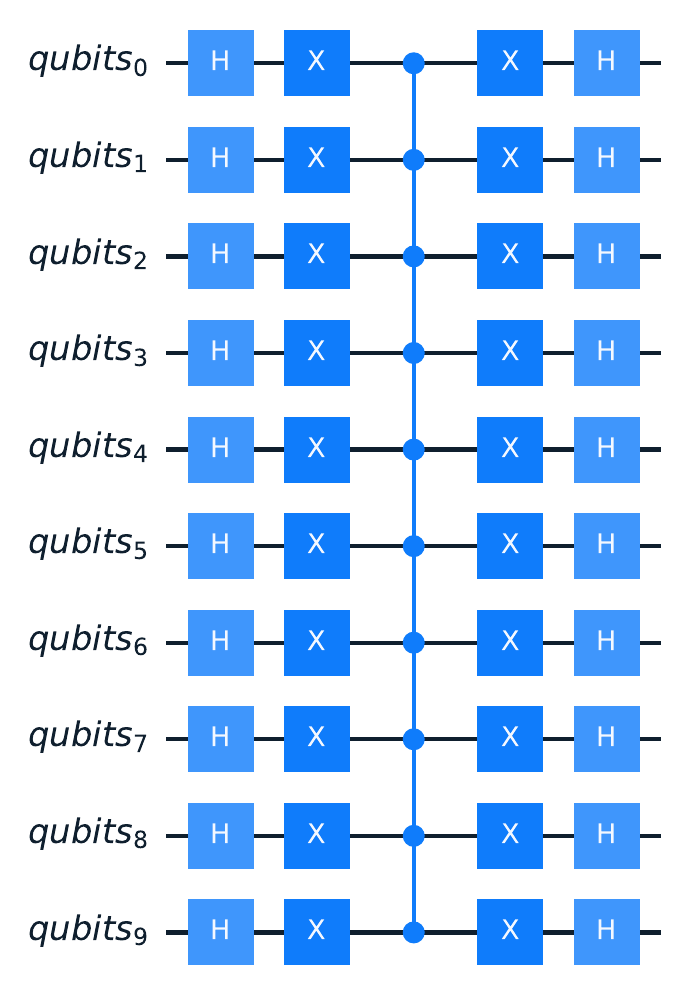}
        \caption{Initial high-level 10-qubit circuit.}
        \label{fig:begin:ex}
    \end{subfigure}
    \caption{Overview of the compilation example setup. (\ref{sub@fig:ket:diffusion}) The hardware-agnostic Ket source code. (\ref{sub@fig:qpu:ex}) The target hardware connectivity. (\ref{sub@fig:begin:ex}) The high-level circuit generated for a 10-qubit system, which serves as the input to the compiler.}\label{fig:ex}
\end{figure}

In this section, we present a concrete example of the quantum compilation process. We will compile the Ket program for the Grover diffusion operator (Figure~\ref{fig:ket:diffusion}) for a target QPU with the connectivity graph depicted in Figure~\ref{fig:qpu:ex}. This hardware, inspired by the IQM Garnet quantum computer, has a native gate set consisting of only $CZ$, $R_Z$, and $\sqrt{X}$ ($\equiv R_X(\tfrac{\pi}{2})$) gates.

Note that the Ket code is written in a qubit-agnostic manner; the specific number of qubits is a classical parameter that must be defined before compilation. For this example, we set the number of qubits to 10. This generates the high-level quantum circuit shown in Figure~\ref{fig:begin:ex}, which serves as the input for our compilation workflow. The subsequent steps of this process are detailed in the remainder of this section.

\subsection{Quantum Gate Decomposition}\label{subsec:decompose}

Despite the ability to apply controls to any gate-like callable in Ket, any multi-controlled gate is ultimately a multi-control, single-target gate. This limits the decomposition algorithms required by the compilation process to a finite set of efficient methods~\cite{Rosa2025}. In Ket, decomposition algorithms are only required for multi-controlled Pauli gates, rotation gates ($SU(2)$ gates), Phase gates, and the Hadamard gate. In practice, this represents a small number of distinct gate classes.

Despite the small number of gate classes, a quantum compiler should implement different decomposition algorithms for the same class, as these algorithms have different trade-offs. Key metrics include the resulting circuit depth, the number of CNOTs required, and the need for ancillary qubits. For example, a highly efficient decomposition for multi-controlled Pauli gates results in a circuit with a linear number of CNOT gates and logarithmic depth relative to the number of controls. However, it requires roughly the same number of ancillary qubits as control qubits~\cite{maslovAdvantagesUsingRelativephase2016,Rosa2025}, which may not be available. In such cases, an alternative decomposition algorithm is needed. The choice of the next-best option depends on whether circuit depth or CNOT count should be prioritized.

The choice of which decomposition algorithm to use in a given situation is made by the compiler; however, it is important for the programmer to be aware of the available algorithms and their associated complexities, as this can influence program performance. For example, although a single-qubit Phase gate is equivalent to a Z-rotation gate, this is not true for their controlled versions. The decomposition of multi-controlled Phase gates requires more resources than that of multi-controlled Z-rotation gates. Therefore, whenever possible, the use of Z-rotation gates is preferable to Phase gates in controlled operations.

In the current NISQ era, with its limited resources, a programmer may be tempted to break abstraction and force the use of a particular decomposition algorithm. Although statically defining the decomposition can result in better performance on certain quantum computers, it may not be optimal on others, making the code's performance hardware-dependent. Conversely, if the compiler dynamically chooses the decomposition algorithm, it can select the appropriate optimization for different architectures. Furthermore, as new and better decomposition algorithms are developed and added to the compiler's pool, code performance can improve over time without manual changes. Code with a statically chosen algorithm will not benefit from these future compiler improvements.

Continuing with our compilation example, the first step is gate decomposition. While knowledge of the target QPU is not strictly necessary at this stage, it allows the compiler to select more efficient decompositions.

\begin{figure}[htbp]
    \centering
    \begin{subfigure}[b]{\linewidth}
        \centering
        \includegraphics[width=\linewidth]{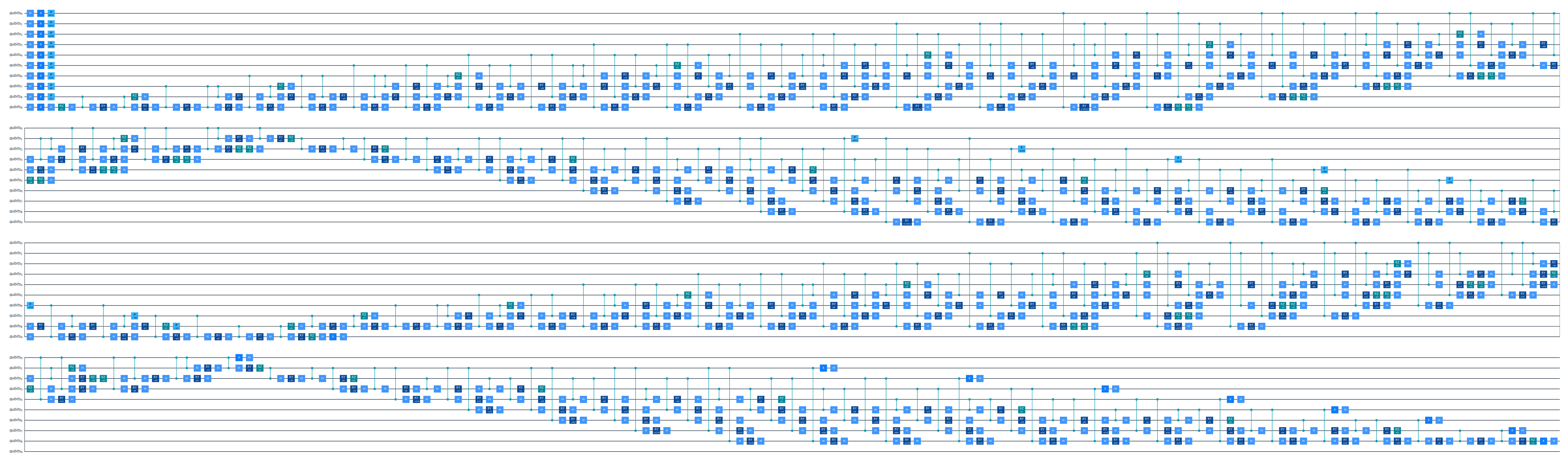}
        \caption{Generic decomposition without using auxiliary qubits.}
        \label{fig:ex:decom1}
    \end{subfigure}
    \begin{subfigure}[b]{\linewidth}
        \centering
        \includegraphics[width=\linewidth]{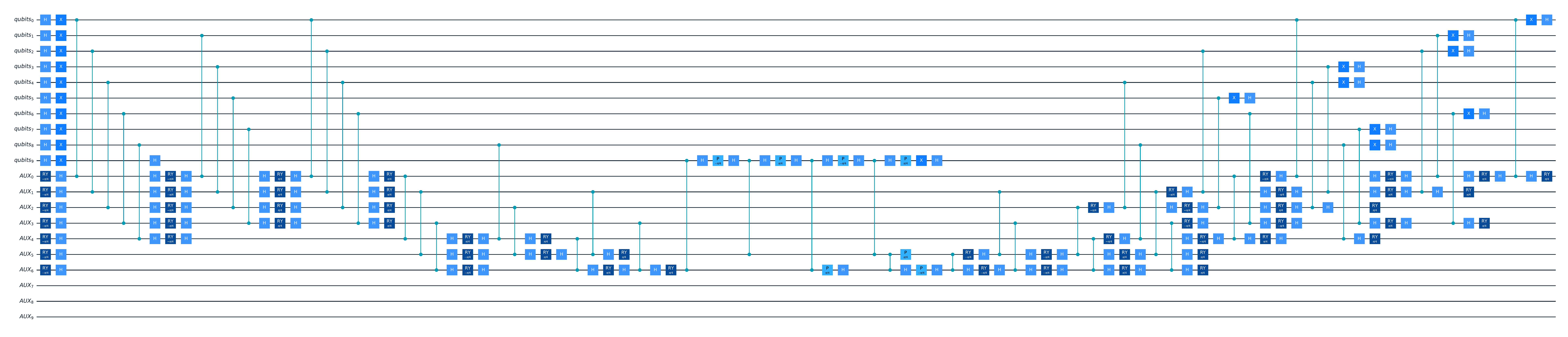}
        \caption{Optimized decomposition using 10 auxiliary qubits available on the target QPU.}
        \label{fig:ex:decom2}
    \end{subfigure}
    \caption{Effect of hardware awareness on gate decomposition. The compiler selects a more efficient strategy when it can use available hardware qubits as auxiliary resources.}
    \label{fig:ex:decom}
\end{figure}

If the compiler has no information about the available hardware qubits, it must use a generic, resource-agnostic algorithm. For the multi-controlled $Z$ gate in our example, this corresponds to a linear-depth decomposition that requires a quadratic number of two-qubit gates~\cite{dasilvaLineardepthQuantumCircuits2022}, as shown in Figure~\ref{fig:ex:decom1}.

However, when the compiler is aware of the target hardware--in this case, a 20-qubit QPU with 10 qubits currently unused--it can leverage these free qubits as auxiliary resources. This enables a more efficient decomposition that reduces the total gate count, as shown in Figure~\ref{fig:ex:decom2}. During this stage, the Ket compiler also translates all two-qubit operations into the target's native gate set; therefore, both circuits are expressed using only CZ and one-qubit gates.

\subsection{Quantum Circuit Mapping}\label{subsec:mapping}

As compilation progresses, the quantum circuit becomes more hardware-specific. Although the use of only single- and two-qubit gates is a consensus in quantum computer design, the circuit mapping stage is highly dependent on the device's architecture. In this step of the compilation, the logical qubits--those allocated by the \texttt{Process} in the high-level program--are mapped to physical qubits. This mapping must respect the qubit connectivity of the QPU. Most often, this process also requires inserting SWAP gates to enable operations between qubits that are not directly connected, which can cause the effective position of a logical qubit to move around the QPU during computation.

Finding the optimal placement of SWAP gates is a computationally difficult problem that requires exponential time to solve exactly\cite{siraichiQubitAllocation2018}; therefore, circuit mapping algorithms rely on heuristics. The state-of-the-art in circuit mapping is based on the SWAP-based Bidirectional (SABRE)~\cite{liTacklingQubitMapping2019} and Dynamic Look-Ahead (DL) heuristics~\cite{zhuDynamicLookAheadHeuristic2020}, both of which take the coupling graph of the QPU as input.

The quality of the final circuit is highly dependent on the initial mapping (the first assignment of logical-to-physical qubits). Since finding the optimal initial mapping is also a hard problem, heuristics are used to find a good starting point~\cite{chowdhuryQubitAllocationStrategies2024}. Many proposed heuristics rely on non-deterministic solvers, which can result in different mappings even for the same input circuit. The DL mapping algorithm, however, proposes deterministic heuristics for the initial mapping. Ket uses the Dynamic Look-Ahead mapping algorithm, as its deterministic nature aids in result reproducibility and helps in identifying performance improvements.

One strategy to improve the initial mapping is to use a forward-and-backward pass approach. The final mapping of a circuit (the position of logical qubits at the end) can be used as the initial mapping for the inverse of that circuit. The final mapping of this inverse circuit can then be used as a new, potentially better, initial mapping for the original circuit. This process can be iterated multiple times, with the best-performing initial mapping chosen from all passes.

Since not all qubits are created equal--the quality of qubits and their connections can vary--some techniques use calibration data to improve the final circuit. This optimization aims not just to reduce the number of SWAPs but to reduce the overall noise during circuit execution~\cite{niuHardwareAwareHeuristicQubit2020}. For example, qubits and connections with smaller error rates can be prioritized. For this purpose, single-qubit error rates are incorporated as weights on the nodes of the coupling graph, and two-qubit error rates are incorporated as weights on the edges.

\begin{figure}[htbp]
    \centering
    \includegraphics[width=\linewidth]{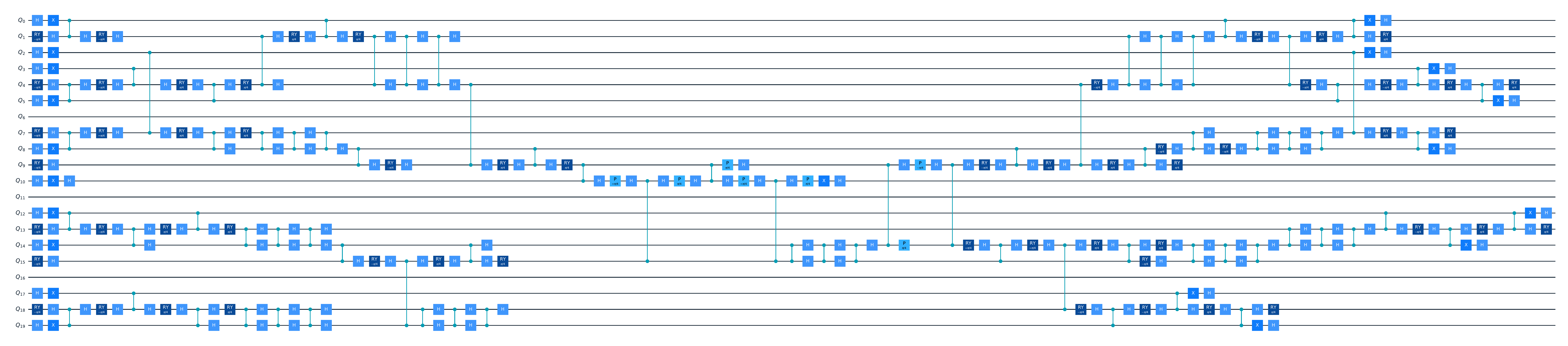}
    \caption{The circuit after the qubit mapping stage. Logical qubits have been assigned to physical qubits.}
    \label{fig:ex:mapping}
\end{figure}

Figure~\ref{fig:ex:mapping} shows the circuit from Figure~\ref{fig:ex:decom2} after circuit mapping, where all two-qubit gates now act exclusively between neighboring physical qubits, respecting the hardware's connectivity graph of Figure~\ref{fig:qpu:ex}. In this example, each SWAP gate is decomposed into three native $CZ$ gates and Hadamard gates.

\subsection{Native Gate Translation}
\label{subsub:native-gate}

Once the circuit has been mapped to the device topology and decomposed into a basis of single- and two-qubit gates (\textit{e.g.}, arbitrary U(2) and CNOT), the final compilation stage is to translate these gates into the native gate set supported by the quantum hardware. Up until this point, the compiler often works with a convenient intermediate representation, such as arbitrary single-qubit unitary gates and CNOTs. However, due to physical implementation and calibration constraints, real quantum computers typically only support a limited, discrete set of single-qubit gates and one or two specific two-qubit entangling gates, which may not necessarily be the CNOT gate (\textit{e.g.}, it could be an iSWAP or a controlled-Z gate). The physical aspects related to calibrating a gate set are addressed in Section~\ref{sec:pulse}.

The native gate translation step does not strictly need to be separate from the preceding compilation stages like gate decomposition and circuit mapping. However, performing this translation too early in the process can unnecessarily increase the complexity of those prior steps. For instance, mapping and optimization algorithms are often simpler to design and more effective when they can operate on a standardized intermediate gate set. In Ket, a mixed approach is taken: the translation of two-qubit gates may be handled during earlier steps, but the final translation of single-qubit gates into the native basis is typically performed as the last step of the decomposition process. A key optimization at this stage is to synthesize sequences of single-qubit gates. Instead of translating each gate in a sequence individually, their corresponding unitary matrices are first multiplied into a single arbitrary unitary. This resultant unitary is then synthesized into an optimal sequence of native gates, which often reduces the overall gate count and execution time.

After the gates are translated into the native set, the circuit is ready for execution. On a quantum simulator, this might be the final step. On physical quantum hardware, however, this native gate circuit is then passed to a pulse scheduler, which generates the precise sequence of analog control pulses (\textit{e.g.}, microwave or laser pulses) that physically manipulate the qubits. Since pulse scheduling depends not only on the hardware's architecture but also on its calibration data (which can vary over time), this final step is typically performed by the hardware provider's control software.

\begin{figure}[htbp]
    \centering
    \includegraphics[width=\linewidth]{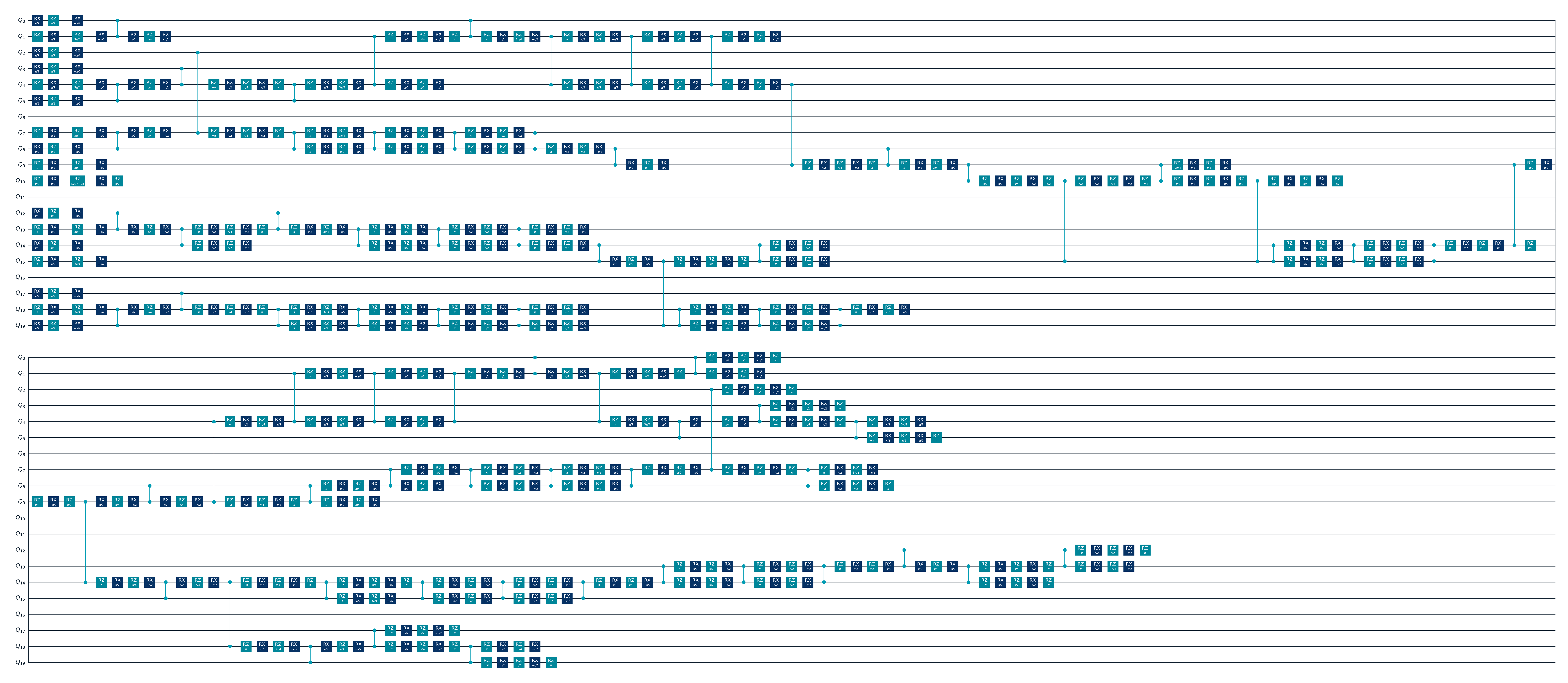}
    \caption{The final compiled circuit for the 10-qubit Grover diffusion operator. All logical operations have been translated into the native hardware gate set ($CZ$, $R_Z$, and $R_X(\tfrac{\pm\pi}{2})$), and all two-qubit gates respect the physical qubit connectivity.}    \label{fig:ex:final}
\end{figure}

To conclude the compilation example, the final stage involves translating all remaining single-qubit gates into the native instruction set of the target hardware. The circuit from the previous step (Figure~\ref{fig:ex:mapping}) is processed, and each single-qubit gate is decomposed into sequences of $R_Z$ and $R_X(\tfrac{\pm\pi}{2})$ gates.
The final, fully compiled circuit is shown in Figure~\ref{fig:ex:final}. It is now expressed entirely in terms of the QPU's native gates and is ready for execution.

\subsection{Expectation Value Engine}
\label{subsub:exp-value}

When the desired result of a quantum computation is a one-shot measurement or a statistical sample of the final state, the compilation steps described thus far are sufficient to prepare the program for execution. However, when the goal is to compute the expectation value of a Hamiltonian ($\langle H \rangle$), as is common in variational algorithms, additional steps managed by the Expectation Value Engine are required.

Naively, to estimate the expectation value of a Hamiltonian expressed as a weighted sum of Pauli strings (\textit{e.g.}, $H = \sum_i c_i P_i$), one would need to execute a different circuit for each term $P_i$ in the sum. This involves adding basis-change rotations at the end of the original circuit to align the measurement basis with the Pauli operators in the term before measuring the qubits. As the number of terms in a Hamiltonian can grow significantly with the problem size, this approach can be prohibitively expensive.

To mitigate this, the engine employs various measurement strategies to reduce the number of distinct circuit executions needed. These techniques rely on identifying and grouping observables that can be measured simultaneously. Prominent strategies include:
\begin{itemize}
    \item \emph{Qubit-wise Commuting} (QWC)~\cite{Verteletskyi2020}: Groups Pauli strings that are composed of commuting Pauli operators on a qubit-by-qubit basis.
    \item \emph{General Commuting}~\cite{Yen2020}: Identifies larger sets of mutually commuting Pauli strings, allowing more terms to be measured from a single circuit execution, though finding optimal groupings can be computationally intensive.
    \item \emph{Classical Shadows}~\cite{Huang2020,Bertuzzi2025}: A randomized measurement technique that constructs a concise classical description of the quantum state from a small number of measurements. This ``shadow'' can then be used to estimate the expectation values of many different observables in classical post-processing, drastically reducing the required number of quantum measurements for large Hamiltonians.
\end{itemize}
Each technique presents a trade-off between the number of quantum circuit executions and the amount of classical post-processing required. The Expectation Value Engine is responsible for executing a measurement strategy by: generating the necessary measurement circuits and reconstructing the final expectation value from the measurement outcomes. For efficiency, the engine typically starts from the circuit produced by the mapping stage, adds the required basis-change gates for a group of measurements, and then sends these new circuits to the final native gate translation stage.

Furthermore, variational quantum algorithms and quantum machine learning models benefit from calculating the gradient of expectation values with respect to parameterized gates. The engine also manages this process. Instead of relying on backpropagation, which is infeasible on quantum hardware, techniques like the parameter-shift rule~\cite{Mitarai2018,Schuld2019} or stochastic parameter-shift~\cite{Banchi2021} are used. These methods also require executing circuit variations and are managed by the engine. The parameter-shift rule provides a strong motivation for delaying native gate translation: it is far simpler to modify a single parameter $\theta$ in an abstract gate like RZ($\theta$) than to track how that change propagates through a sequence of decomposed fixed native gates.

\subsection{Circuit Optimization}
\label{subsub:optimization}

Quantum circuit optimization is not a single compilation stage but a transversal process that can be performed at multiple layers of the software stack. The goal is to reduce circuit resources (such as gate count, depth, or CNOT count) to improve execution fidelity on noisy hardware. The available optimization techniques depend on the circuit representation at each layer.

Optimization can range from high-level logical reductions to low-level hardware-aware passes. For instance, at the top of the stack, high-level programming constructs can be used to reduce the number of controlled operations~\cite{rosaOptimizingGateDecomposition2025}. During compilation, common techniques include canceling adjacent inverse gates ($U U^\dagger \to I$) and merging consecutive single-qubit rotations. More advanced and computationally intensive techniques, such as resynthesizing entire circuit blocks using methods like the ZX-calculus~\cite{Coecke2011}, can provide significant reductions in gate count and are an active area of research.

\section{Physical Realization of Quantum Programs}\label{sec:pulse}

The quantum compilation process, detailed in Section~\ref{sec:compiler}, transforms a high-level quantum program into a circuit composed of native gates that respects the hardware's physical topology. However, this gate-based description is still an abstraction. A Quantum Processing Unit (QPU) does not execute gates directly; rather, it is a physical system manipulated by precisely timed analog control signals. The final step in running a quantum program is therefore to translate the sequence of native gates into a schedule of control pulses--typically microwave or laser pulses--that are sent to the QPU by classical control hardware.

This translation from the digital abstraction of gates to the analog reality of pulses is made possible by the careful definition and calibration of a native gate set. Each gate in this set, such as the $CZ$, $R_Z$, and $\sqrt{X}$ gates from our previous example, corresponds to a pre-characterized pulse shape, frequency, and duration that implements the desired unitary transformation. This section describes what is required at the hardware level to define, calibrate, and execute these pulse-level instructions.

While several technologies exist for building quantum computers, including trapped ions and photonics, we focus here on \emph{superconducting transmon qubits}, one of the most prominent platforms~\cite{Maslov2019, Gill2022}. These qubits are engineered electrical circuits composed of capacitors, inductors, and Josephson junctions. The physics of superconductivity--where electrons form ``Cooper pairs'' and flow without resistance below a critical temperature--allows these circuits to maintain quantum coherence for long periods~\cite[Chapter 1]{Fossheim2004}. The Josephson junction provides the necessary non-linearity, creating unequally spaced quantized energy levels. The lowest two levels are chosen to encode the qubit states $\ket{0}$ and $\ket{1}$, which can then be manipulated by applying electromagnetic pulses at the appropriate transition frequency.

In this section, we describe how these superconducting qubits are programmed and manipulated at the pulse level, which constitutes the lowest layer of the quantum software stack. We will cover the calibration procedures required to characterize the hardware and ensure high-fidelity operations. We focus on the principles of quantum control, abstracting away details of low-level circuit design; for a more complete treatment of the physical implementation, we refer the reader to~\cite{Krantz2019} and~\cite{Blais2021}. We begin by introducing the physical principles behind qubit control (Section~\ref{subsec:pulse-modeling}), then present the essential calibration routines for single-qubit gates (Section~\ref{subsec:single-qubit-gates-calibration}), multi-qubit gates (Section~\ref{subsec:multi-qubit-gates-calibration}), and measurement (Section~\ref{subsec:measurement-calibration}).

\subsection{Control Pulse Theory}\label{subsec:pulse-modeling}

As briefly introduced before, quantum gates are implemented using microwave pulses. From quantum mechanics, we know that a two-level quantum system undergoes coherent oscillations between its basis states, $\ket{0}$ and $\ket{1}$, when driven by a resonant oscillatory field~\cite[Section 7.5.3]{nielsenQuantumComputationQuantum2010}. This coherent population transfer is known as Rabi oscillation. Constructing quantum gates essentially means precisely controlling the Rabi frequency (which sets how fast the state oscillates between $\ket{0}$ and $\ket{1}$) and the rotation axis on the Bloch sphere, e.g., $x$ or $y$. By adjusting these parameters and the pulse duration, we steer the quantum state to the desired point.

\begin{figure}[h]
    \centering
    \includegraphics[width=0.45\textwidth]{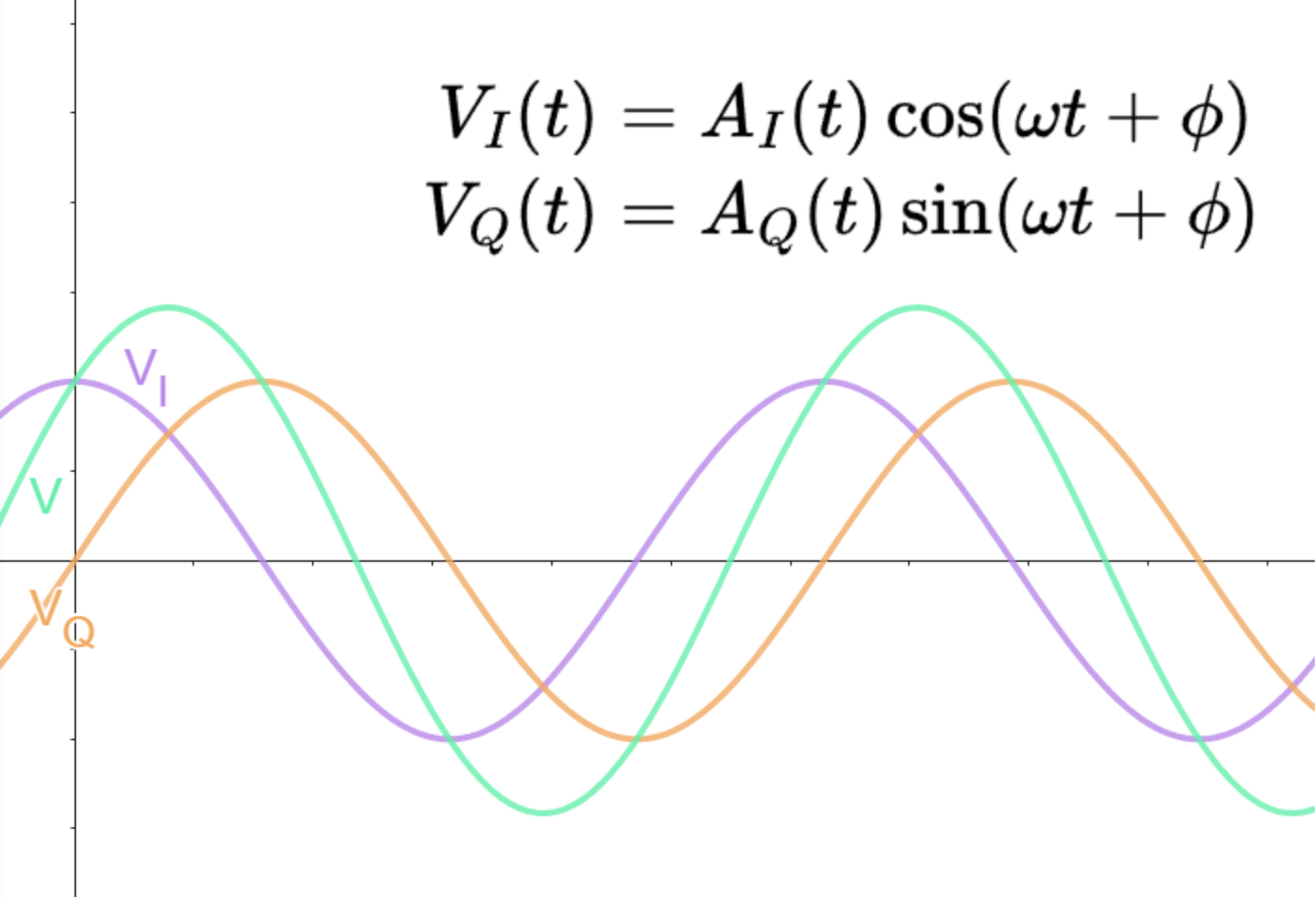}
    \caption{Illustration of $IQ$ modulation. The in-phase ($I$) and quadrature ($Q$) components, shifted by $\pi/2$, combine to form a pulse whose amplitude and phase are determined by their relative contributions.}
    \label{fig:IQ-modulator}
\end{figure}

A pulse is characterized by a waveform, typically represented by sine or cosine functions. In this context, we commonly utilize $IQ$ modulation, which involves two distinct pulses: the in-phase ($I$) and the quadrature ($Q$), with a phase shift of $\pi/2$. The final pulse is obtained by combining the $I$ and $Q$ components, which results in a pulse with an amplitude and phase that depends on the two components. This can be visualized in Figure \ref{fig:IQ-modulator}, where $\omega$ represents the frequency of the pulse, $A(t)=A_I(t)+iA_Q(t)$ denotes the dimensionless complex envelope (amplitude modulation), $\phi$ is the phase, $t$ is time, and $V(t)=V_I(t) + V_Q(t)$ is the time-dependent voltage signal applied to the system.

When building control pulses, we modulate the attributes of the control pulse. The frequency $\omega$ is typically set to the qubit's resonance frequency to drive coherent $\ket{0}\leftrightarrow\ket{1}$ oscillations. The phase $\phi$ determines the rotation axis of the Rabi oscillation on the Bloch sphere. The amplitude $A$ values indicate how fast the quantum state oscillates: the higher the value, the faster the quantum state oscillates. Therefore, to build faster gates, higher amplitude values are needed, and for slower gates, lower amplitude values suffice.

\subsubsection{Control Hamiltonian}

The complex envelope $A(t)$ defines the in-phase and quadrature components of the pulse as programmed in the control system. These are dimensionless quantities that specify the analog waveform to be delivered by the control electronics, typically constrained within a unit disk, $\left|A(t)\right| \leq 1$. However, the actual interaction strength between the control field and the qubit is described by the Hamiltonian (adapted from~\cite{Carvalho2021}):
\begin{equation} \label{eq:hamiltonian-pulse-shape}
    H(t) = \frac{1}{2} \left(I(t)\sigma_x + Q(t)\sigma_y\right),
\end{equation}
where $\gamma(t)=I(t) + iQ(t)$ is the control pulse waveform in Rabi frequency units. The physical quantities $I$ and $Q$ are proportional -- though not exactly linear -- to the voltages delivered to the qubit and determine the rotation rate and axis of the state on the Bloch sphere. In this representation, the $I$ component drives rotations around the $x$-axis of the Bloch sphere, while the $Q$ component drives rotations around the $y$-axis. Intermediate phase values drive rotations around other axes in the $xy$-plane.

It is possible to determine the rotation angle induced by a control pulse by integrating the control Hamiltonian over time. For an in-phase control pulse (with $Q(t) = 0$), the rotation angle $\Theta$ around $x$-axis is given by~\cite[Section IV D]{Krantz2019}:
\begin{equation}
    \Theta=\int_0^t I(t') dt'.
\end{equation}
This means that, for example, to implement a $\pi$-pulse around $x$-axis, one must choose the waveform $I(t)$ such that the total area under the pulse satisfies $\Theta = \pi$.

\subsubsection{Waveform Sampling}

In this theoretical model, we define the control pulse as a continuous complex-valued function $A(t)$. This function represents the envelope of the analog voltage signal that drives the qubit. However, in practice, the generation of such pulses is constrained by the limitations of the control hardware. Specifically, the control electronics rely on digital waveform generation, in which the continuous function $A(t)$ is represented by a discrete sequence of samples:
\begin{equation}
    A[n]=A(n.\mathit{dt})
\end{equation}
where $\mathit{dt}$ is the sampling time, a fixed temporal resolution determined by the sampling rate of the control system (\textit{e.g.}, $\SI{1}{\giga S / \second} = \SI{1}{\nano\second}$). This means that the pulse envelope is not specified as a smooth, continuous curve, but rather as a list of values at fixed intervals.

The discrete samples are then sent to a Digital-to-Analog Converter (DAC)~\cite{DAC}, which transforms the digital sequence into an analog voltage signal $V(t)$. This analog signal is what is physically transmitted through the control line to the qubit. However, the DAC output is not a perfect stepwise voltage; instead, the actual signal $V(t)$ is an approximation of the ideal piecewise-constant waveform implied by the digital samples. From a signal processing perspective, the amplitude modulation of this signal can be analyzed using the Fourier transform, which reveals that the control pulse may contain a broad spectrum of frequency components beyond the target frequency $\omega$.

An intuitive analogy can be made with musical instruments, such as a guitar. When a string is plucked to produce a note at a certain frequency, the sound emitted is not a pure sine wave, but rather a combination of multiple frequencies due to the finite duration and mechanical properties of the excitation. These additional frequency components may resonate with other strings, causing them to vibrate sympathetically even if they were not directly plucked. Similarly, in a quantum processor, a control pulse aimed at a specific qubit transition may unintentionally contain spectral components that resonate with other transitions--either within the same qubit (causing leakage to higher levels) or in neighboring qubits (leading to crosstalk).

Because hardware cannot perfectly reproduce pulses with abrupt transitions or extremely wide bandwidths, control sequences in experiments are often designed with these physical limitations in mind.

\subsubsection{Pulse Shaping}

The amplitude modulation of a control pulse plays a crucial role in determining the fidelity of quantum operations. The goal of pulse shaping is to modulate the amplitude in a way that enhances gate fidelity and increases robustness against various sources of error. There are three main approaches to pulse design: analytical techniques, quantum optimal control, and reinforcement learning.

In analytical techniques, optimal pulse shapes are derived based on a complete geometric understanding of the system dynamics~\cite{Koch2022}. These pulses are described by mathematical expressions with tunable parameters. Quantum optimal control, on the other hand, formulates pulse shaping as a numerical optimization problem. In reinforcement learning, control pulses are learned directly from interactions with the quantum hardware.

Here we describe some common pulse shaping methods utilized in these approaches.

\paragraph*{Basic Pulse Shapes}

\begin{figure}[h]
    \centering
    \includegraphics[width=0.9\linewidth]{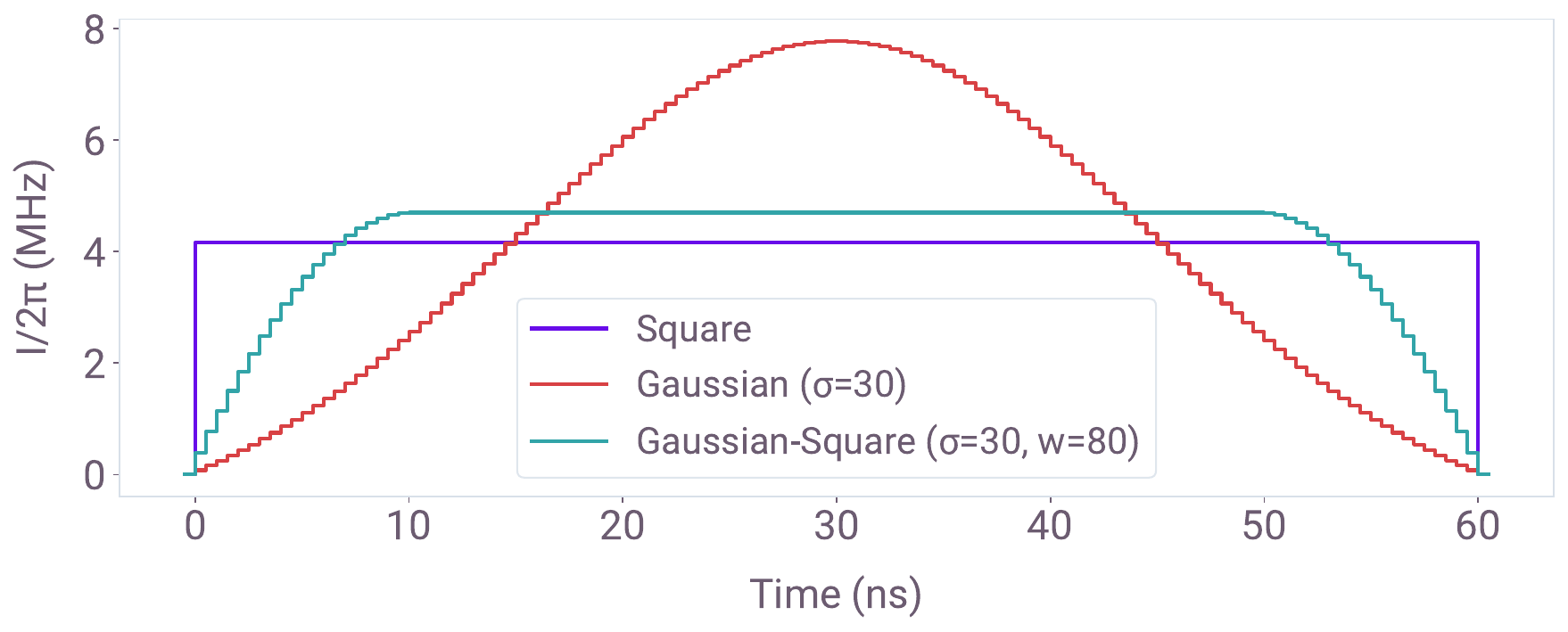}
    \caption{Illustration of three basic pulse shapes in the in-phase (I) quadrature: square (purple), Gaussian (red), and Gaussian-square (green). Each pulse implements a $\pi/2$ rotation with a total duration of \SI{60}{\nano\second} and a sampling time of \SI{0.5}{\nano\second}.}
    \label{fig:basic-pulse-shapes}
\end{figure}

The square pulse, shown in purple in Fig. \ref{fig:basic-pulse-shapes}, represents a control signal whose amplitude remains constant throughout its duration. Despite its simplicity, the square pulse has very sharp rise and fall edges, which introduce unwanted high-frequency content when analyzed in the frequency domain. These high frequencies can cause leakage errors, where the qubit unintentionally transitions to states beyond $\ket{1}$, such as $\ket{2}$ or $\ket{3}$. This happens because some of the frequency components of the pulse may match the transition frequencies to those higher energy levels, and can do so with significant strength. In addition, ideal square pulses cannot be physically implemented because creating perfectly sharp edges would require changing the signal infinitely fast--something that is not possible with real electronic equipment.

The effect of a square pulse on a qubit is straightforward to analyze, as it corresponds to the area under a constant amplitude waveform--\textit{i.e.}, a rectangle. For example, to implement a $\pi/2$ rotation over a duration of \SI{60}{\nano\second}, we can compute the required Rabi frequency as $\frac{\pi/2}{\SI{60}{\nano\second}} \approx \SI{26\,179\,938}{\radian\per\second} \approx \SI{4.167}{\mega\hertz}$. The Rabi frequency defines the rate at which the qubit state oscillates between $\ket{0}$ and $\ket{1}$ and is directly related to the pulse amplitude. To determine the amplitude that produces this target Rabi frequency, one typically performs a Rabi oscillation experiment and may need to fine-tune the amplitude-to-frequency mapping. This calibration process is discussed in detail in Section~\ref{subsec:single-qubit-gates-calibration}. The rotation axis on the Bloch sphere is controlled by the phase of the pulse, which in turn depends on the in-phase ($I$) and quadrature ($Q$) components: applying the pulse through the $I$ channel yields an $x$ rotation, while using the $Q$ channel results in a $y$ rotation.

To suppress the high-frequency components of a square pulse, Gaussian-shaped envelopes are commonly used (see Fig.~\ref{fig:basic-pulse-shapes}). For the Gaussian-based pulse shapes considered below, we define a normalization function
\begin{equation}
    \mathcal{N}[g](t) = A \,\frac{g(t) - g(-1)}{1 - g(-1)} \,,
\end{equation}
where $g(t)$ is the unnormalized envelope and $A$ is the amplitude scaling factor. The subtraction at $t=-1$ ensures that $\mathcal{N}[g](t)$ vanishes exactly one time step before ($t = -1\,dt$) and after ($t = d + 1\,dt$) the pulse duration.

The Gaussian envelope is defined as
\begin{equation}\label{eq:gaussian}
    g_{\text{gauss}}(t) = \exp\!\left[-\frac{(t - d/2)^2}{2\sigma^2}\right] \,,
\end{equation}
and the normalized waveform is simply $f_\text{gauss}(t) = \mathcal{N}[g_\text{gauss}](t) \,$. Here, $d$ is the total pulse duration and $\sigma$ is the standard deviation controlling the smoothness of the rise and fall of the pulse.

While Gaussian pulses offer smoother transitions and attenuate some high-frequency components caused by abrupt amplitude changes, maintaining the same area under the curve--or equivalently, the same rotation--requires higher amplitude values near the center of the Gaussian shape. When implementing faster Gaussian pulses, the issue of high-frequency components arise again, similarly to square pulses, but now it is a consequence of the pulse shape itself rather than its implementation.

An intermediary between Gaussian and square pulses is the Gaussian-square pulse (also shown in Fig.~\ref{fig:basic-pulse-shapes}). This pulse shape combines Gaussian rise and fall sections with a square (constant) plateau at its center. The unnormalized envelope is given by
\begin{equation}
    g_\text{gsq}(t) =
    \begin{cases}
        \exp\!\left(-\dfrac{(t-r)^2}{2\sigma^2}\right),     & t < r,          \\
        1,                                                  & r \leq t < r+w, \\
        \exp\!\left(-\dfrac{(t-(r+w))^2}{2\sigma^2}\right), & t \geq r+w \,,
    \end{cases}
\end{equation}
and the normalized waveform is $f_\text{gsq}(t) = \mathcal{N}[g_\text{gsq}](t) \,$. Here, $\sigma$ represents the standard deviation of the Gaussian rise and fall, $w$ denotes the width of the plateau, and $r = d - w$ is the risefall factor, with $d$ being the total pulse duration.

\paragraph*{DRAG}

To suppress leakage errors from fast Gaussian pulses, a technique called Derivative Removal by Adiabatic Gate (DRAG)~\cite{Gambetta2009} is often employed for implementing single-qubit gates. This approach involves a pulse shape combining a standard Gaussian envelope in the $I$ component with an additional derivative component in the $Q$ component. The pulse can be written as
\begin{equation}
    f(t) = f_\text{gauss}(t)\,\left[1 - i\,B\,\left(\frac{t-d/2}{\sigma^2}\right)\right],
    \qquad 0 \leq t < d \,,
\end{equation}
$f_\text{gauss}(t)$ is the normalized Gaussian envelope. Here, all parameters retain the same meaning as previously defined, except for $B$, which determines the strength of the $Q$-quadrature correction.

The idea behind the DRAG waveform is to apply an additional quadrature component ($Q$) to the control pulse in order to mitigate leakage into higher energy levels. Without this correction, the $I$ component of the pulse (typically shaped as a Gaussian envelope) exhibits Fourier components at frequencies close to the $\ket{1}\!\to\!\ket{2}$ transition, which can inadvertently drive population out of the computational subspace. By introducing the properly scaled $Q$ component, the DRAG pulse cancels out these unwanted high-frequency contributions, thereby reducing the spectral weight around the leakage transition.

\paragraph*{Quantum Optimal Control}

Quantum optimal control tackles the pulse design problem by directly formulating it as a numerical optimization task. The goal is to find control fields--typically piecewise-constant or smoothly varying waveforms--that drive the quantum system to implement a target unitary operation with high fidelity.

The challenge lies in the high dimensionality and nonlinearity of the control landscape. Control pulses must respect physical constraints, such as bandwidth and amplitude limits, while also compensating for hardware imperfections and environmental noise. Furthermore, the cost functions involved (\textit{e.g.}, gate infidelity) are often highly non-convex, requiring advanced optimization strategies to avoid local minima and ensure convergence to effective solutions.

Within this framework, the Hamiltonian in Equation~\ref{eq:hamiltonian-pulse-shape} serves as the control model, and can be extended to account for error processes. For instance, dephasing noise $\eta(t)$ and amplitude miscalibrations $\beta(t)$ can be incorporated as:
\begin{equation}
    H_{\text{robust}}(t) = (1 + \beta(t))H(t) + \eta(t)\sigma_z,
\end{equation}
which enables the synthesis of control pulses that are intrinsically robust to these types of noise.

Tools such as \textit{Boulder Opal} by Q-CTRL exemplify this approach. By leveraging numerical optimization over realistic noise models~\cite{Carvalho2021}, they produce high-fidelity, noise-robust pulse shapes. More recently, Boulder Opal has integrated reinforcement learning with optimal control, using autonomous agents to discover optimal pulses directly from hardware interaction~\cite{Baum2021}, bridging model-based and data-driven control paradigms.

The techniques discussed so far are necessary for designing high-fidelity waveforms that implement quantum logic gates. However, achieving reliable quantum operations on actual hardware also requires a precise calibration of system parameters and control channels. The next sections present key calibration procedures that enable accurate control of quantum systems and ensure that the pulses designed theoretically can be faithfully implemented on physical devices.

\subsection{Single-Qubit Gates Calibration}\label{subsec:single-qubit-gates-calibration}

A fundamental goal of calibration is to enable the accurate implementation of a set of quantum gates that is universal for quantum computation. In this context, any quantum algorithm can be decomposed into sequences of single-qubit gates and at least one two-qubit entangling gate, such as the $\mathrm{CNOT}$~\cite[Section~4.5.2]{nielsenQuantumComputationQuantum2010}. Various universal gate sets are known, including $\{\mathrm{CNOT}, H, T\}$~\cite{BOYKIN2000101}, $\{\mathrm{CNOT}, R_y(\pi/4), S\}$~\cite{AYuKitaev_1997}, and $\{\mathrm{Toffoli}, H\}$~\cite{10.5555/2011508.2011515}. For example, IBMQ devices adopt a native gate set consisting of $\{\sqrt{X}, R_z, \mathrm{CNOT}\}$.

For a universal set of single-qubit operations, a widely used choice is $\{\sqrt{X}, R_z\}$, motivated by the fact that arbitrary single-qubit unitaries can be efficiently decomposed using only two $\sqrt{X}$ (\textit{i.e.}, $R_x(\pi/2)$) gates and three $R_z$ rotations. As shown in~\cite{McKay2017} and, any unitary operation $U(\theta, \phi, \lambda)$ can be expressed as:
\begin{equation}\label{eq:rzsxrzsxrz}
    U\left(\theta, \phi, \lambda\right)=
    R_z\left(\lambda\right)
    \sqrt{X}
    R_z\left(\theta\right)
    \sqrt{X}^{-1}
    R_z\left(\phi\right).
\end{equation}
This decomposition breaks down an arbitrary single-qubit gate with three Euler angles $U$ into five other quantum gates: two $\sqrt{X}$ (or $R_x(\pi/2)$) gates and three $R_z$ gates. Such decomposition enables the implementation of arbitrary single-qubit gates using just a $\sqrt{X}$ gate, given the virtual implementation of $R_z$ gates (\textit{i.e.}, through phase adjustments in the control system, as discussed in Section \ref{subsubsec:virtual-z-gates}).

An alternative approach involves implementing a $U$ gate using arbitrary $x$ rotations. While this method reduces the number of quantum gates, it relies on the capability to execute arbitrary $x$ rotations:
\begin{equation}\label{eq:rzrxrz}
    U\left(\theta, \phi, \lambda\right)=
    R_z\left(\lambda-\pi/2\right)
    R_x\left(\theta\right)
    R_z\left(\phi-3\pi/2\right).
\end{equation}

The calibration process of quantum gates consists of adjusting the control hardware parameters in order to implement the desired theoretical control pulse. For single-qubit gates, we need to calibrate the $(I, Q) \leftrightarrow (A_I, A_Q)$ mapping--that is, determine the hardware input amplitudes $A_I$ and $A_Q$ that produce the desired Rabi frequencies $I$ and $Q$. In what follows, we describe a typical process for this calibration. In Section~\ref{subsubsec:rabi-experiment}, we introduce a standard Rabi experiment that provides an overview of this mapping. Then, in Section~\ref{subsubsec:fine-tuning}, we show how to fine-tune it. Finally, in Section~\ref{subsubsec:virtual-z-gates}, we explain how virtual $Z$ gates are implemented in practice.

\subsubsection{Rabi Experiment}\label{subsubsec:rabi-experiment}

The first step in constructing quantum gates is to understand how the hardware responds to control pulses. Recall that the complex envelope $A(t)=A_I(t)+iA_Q(t)$ defines the pulse shape as programmed in the control system, while the physical effect on the qubit is governed by the components $I(t)$ and $Q(t)$ in the control Hamiltonian (Equation \ref{eq:hamiltonian-pulse-shape}).

The Rabi experiment consists of calibrating the relation $(I, Q) \leftrightarrow (A_I, A_Q)$ by fixing the frequency and phase of the pulse and sweeping over different envelope amplitudes ($A$). For each amplitude, the qubit is driven with a constant pulse over a range of durations, and the resulting state populations are measured.

\begin{figure}[!b]
    \centering
    \includegraphics[width=0.8\textwidth]{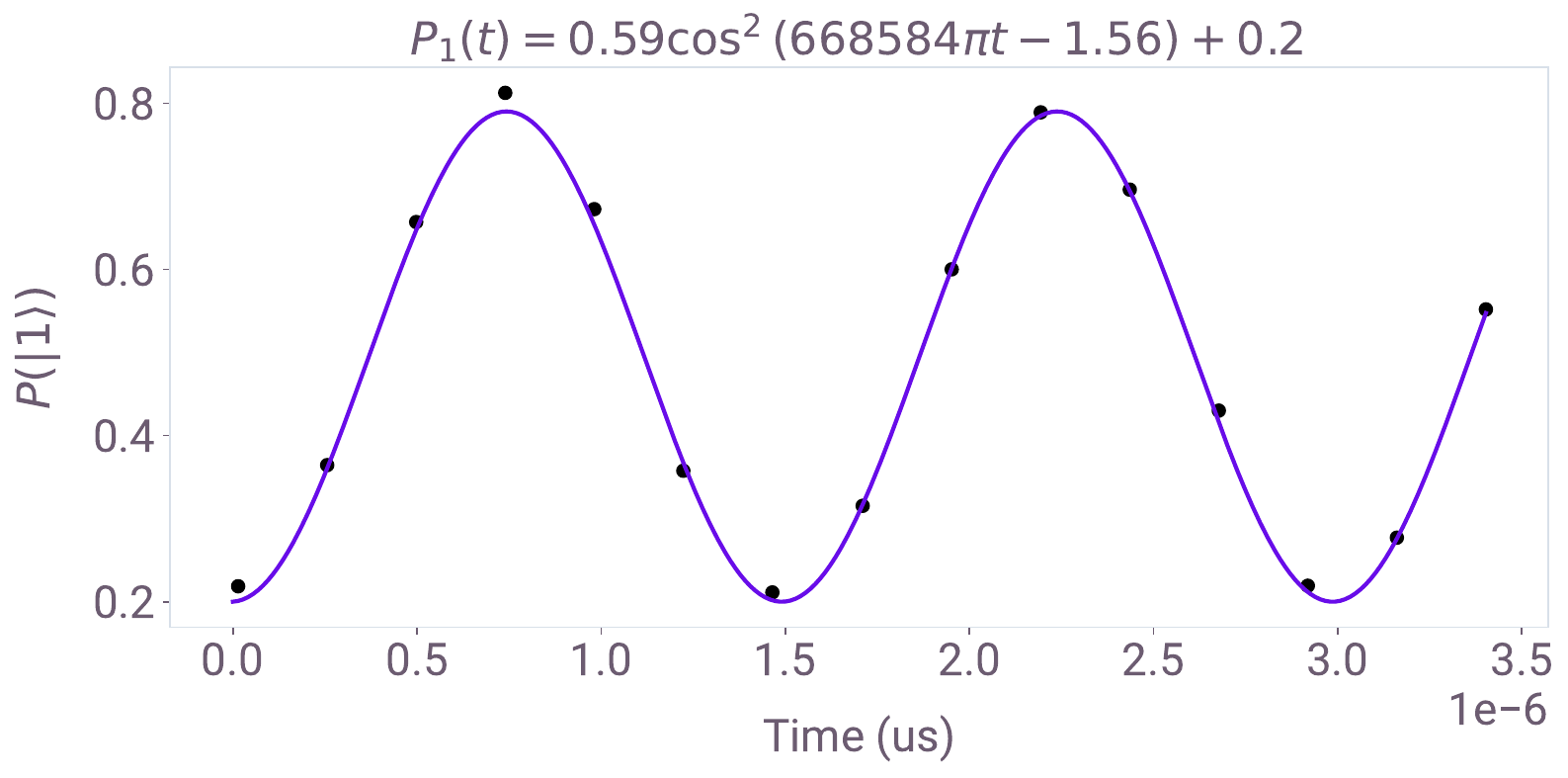}
    \caption{Rabi oscillation obtained from qubit 0 on the IBMQ Kyoto quantum computer using a pulse with an amplitude of $0.005$.}
    \label{fig:rabi-oscillations-exmp}
\end{figure}

For a fixed drive amplitude, the population of the qubit excited state oscillates over time due to Rabi oscillations. This behavior can be modeled by
\begin{equation}
    P_1(t) = \mathcal{A} \cos^2\left(\Omega\pi t + \Phi\right) + \delta,
\end{equation}
where $t$ denotes the evolution time, $\Omega$ is the Rabi frequency scaled to a full rotation ($2\pi$), and $\mathcal{A}$, $\Phi$, and $\delta$ are fitting parameters.

To illustrate, Figure~\ref{fig:rabi-oscillations-exmp} shows the Rabi oscillation obtained from qubit 0 on the IBMQ Kyoto (retired) quantum computer using a pulse with an amplitude of $0.005$. The data points represent the experimental results, while the continuous line corresponds to the fitted function $P_1(t)$. In this specific instance, the extracted Rabi frequency is approximately \SI{668584}{\hertz}, indicating that the qubit state oscillates between $\ket{0}$ and $\ket{1}$ roughly every $1/668584 \approx \SI{1496}{\nano\second}$. Thus, to implement an $X$ gate using a pulse of amplitude $0.005$ on this qubit, a pulse duration of approximately $\SI{1496}{\nano\second}/2 \approx \SI{748}{\nano\second}$ would be required.

\begin{figure}[!h]
    \centering
    \includegraphics[width=0.7\textwidth]{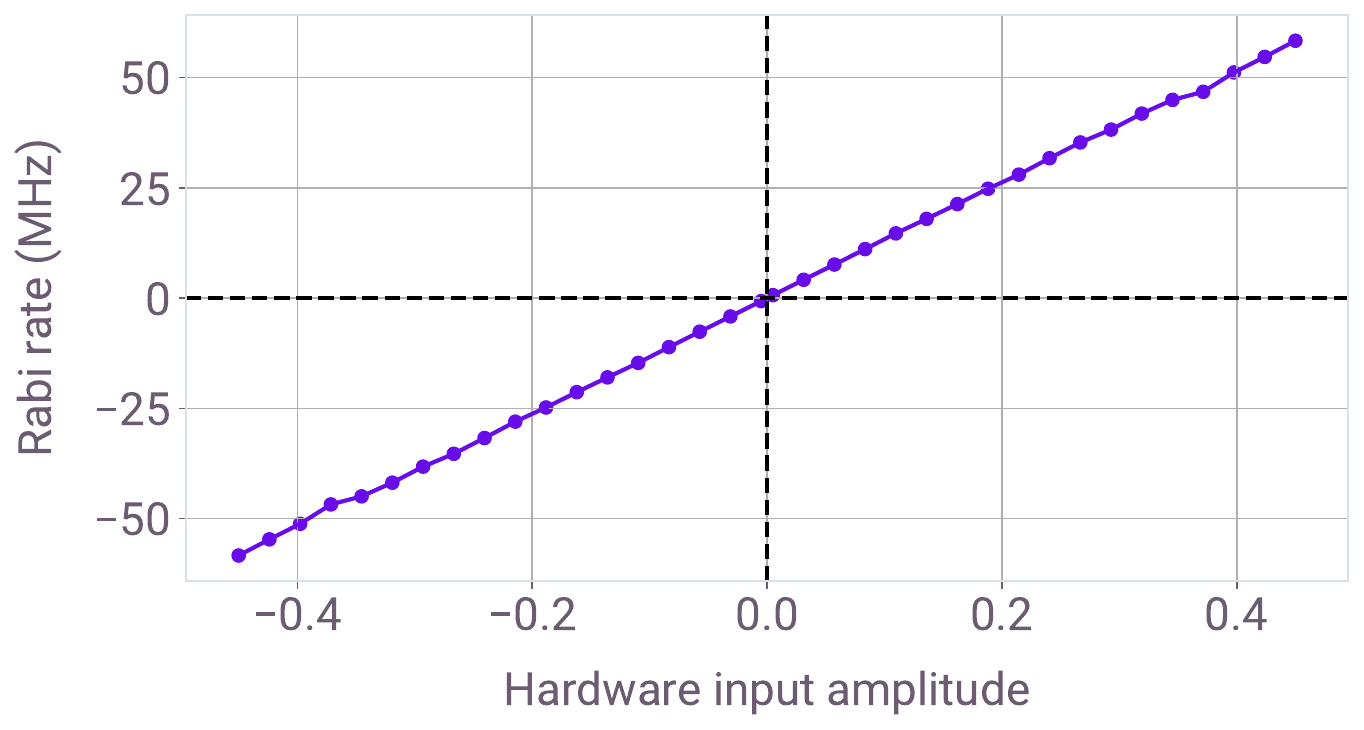}
    \caption{Rabi experiment results conducted on qubit 0 of the IBMQ Kyoto quantum computer.}
    \label{fig:rabi-exmp}
\end{figure}

The function fitting process is repeated for each selected amplitude value to derive the corresponding Rabi frequency $\Omega$. As a result, we establish a correlation between the hardware input amplitude $A$ and the Rabi frequency $\Omega$ by interpolating the experimental pairs $(\Omega, A)$. The final outcome of a Rabi experiment conducted on qubit 0 of Kyoto is shown in Figure~\ref{fig:rabi-exmp}. In this experiment, Rabi frequencies were determined for 18 amplitude values ranging from $0.005$ to $0.45$ (with results mirrored for negative values). This plotted function serves as a fundamental reference for implementing quantum gates, providing the pulse amplitude corresponding to a given Rabi frequency. When designing quantum gates, with the duration and rotation angle defined, we calculate the required Rabi frequency for the given time interval and use this function to obtain the appropriate amplitude.

\subsubsection{Fine-Tuning}\label{subsubsec:fine-tuning}

The interpolation of Rabi experiment data does not perfectly represent the relationship $\Omega \leftrightarrow A$. The exact Rabi frequencies are known only for the specific set of amplitudes used in the experiment. As a result, estimating frequencies for other amplitudes may introduce under- or over-rotations. To improve this mapping, the control pulse requires fine-tuning. This process typically involves sweeping the amplitude around those estimate by the interpolation to identify the amplitude that achieves the highest fidelity\footnote{Fidelity measures the closeness between quantum states. For pure states $\ket{\psi}$ and $\ket{\phi}$, it is defined as $F\left(\ket{\psi},\ket{\phi}\right)=\left|\braket{\psi|\phi}\right|^2$~\cite[Section 9.2.2]{nielsenQuantumComputationQuantum2010}.}. Specifically, the goal is to determine the parameters $S_{amp}$ and $S_{rel}$ in the following equation to maximize fidelity~\cite{Carvalho2021, Baum2021}:
\begin{equation}\label{eq:fine-tuning}
    A_{\text{fine-tuned}}(t)=S_{amp}\left(S_{rel}A_I(t) + iA_Q(t)\right).
\end{equation}

\begin{figure}[!h]
    \centering
    \includegraphics[width=0.8\textwidth]{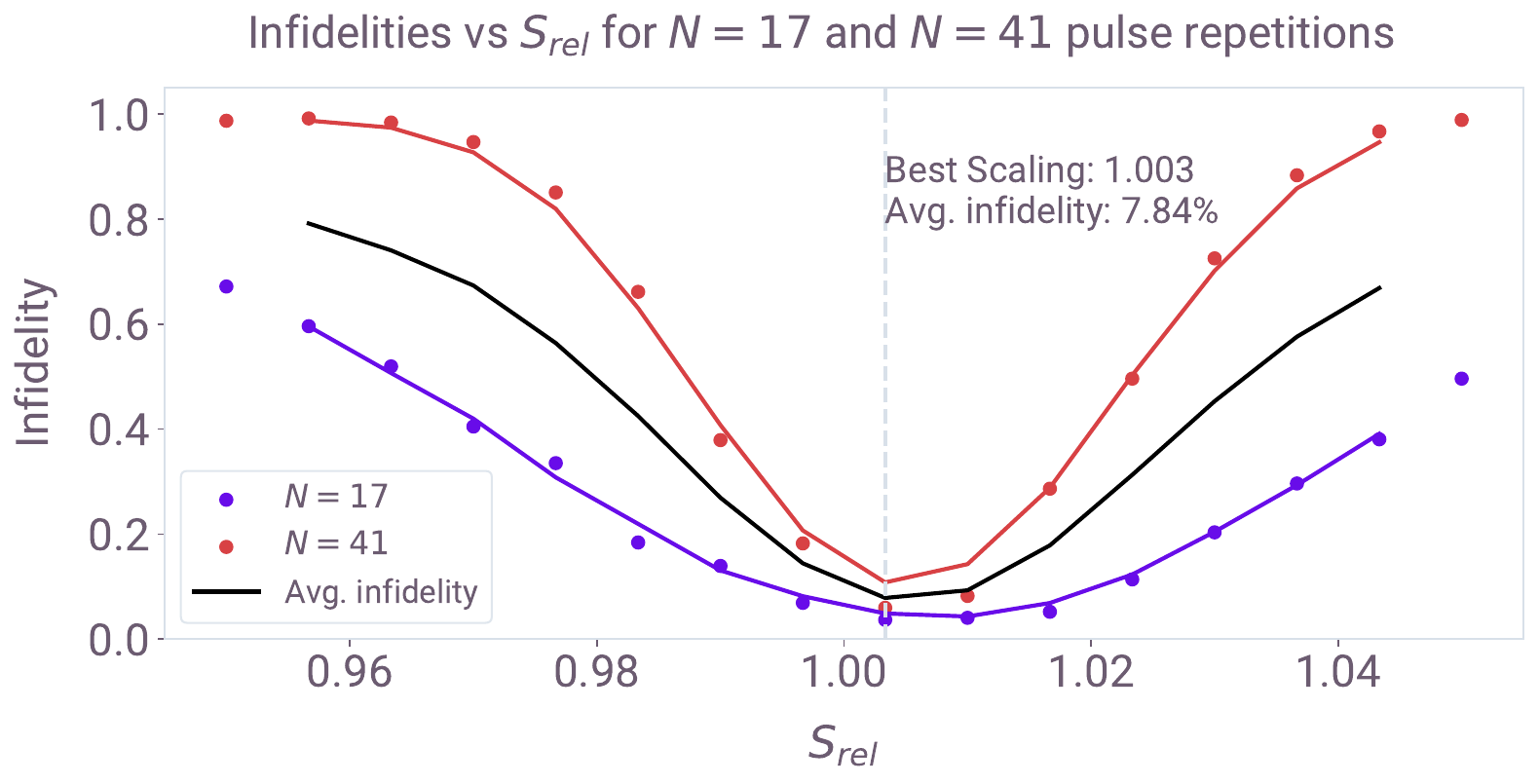}
    \caption{Fine-tuning experiment for a $\pi/2$ pulse on IBMQ Brisbane, using a $\SI{256}{\nano\second}$ Q-CTRL dephasing-robust control. The plot shows the infidelity as a function of the relative scaling $S_{rel}$ for $N=17$ and $N=41$ repetitions. The black curve denotes the averaged infidelity, used to extract the optimal $S_{rel}$.}
    \label{fig:fine-tuning}
\end{figure}

Figure \ref{fig:fine-tuning} presents an example of a fine-tuning experiment. The control pulse amplitude $A$ was swept around the value estimated by the interpolation, and the resulting fidelity was measured for two different pulse repetitions. Repeating the control pulse serves to amplify any systematic errors, making them more observable. The highest measured value of $S_{rel}$ was approximately $1.003$, with an average infidelity of about 7.84\% across the two repetitions.

\subsubsection{Virtual $Z$ Gates}\label{subsubsec:virtual-z-gates}

Unlike $x$ or $y$ rotations, which require the application of shaped microwave pulses in the $xy$-plane, rotations around the $z$ axis can be implemented virtually--without applying any physical pulse. These so-called virtual $Z$ gates are realized by simply updating the phase reference frame of subsequent control pulses. As a result, $z$ rotations have effectively zero duration and do not introduce additional errors or hardware overhead~\cite{McKay2017}.

From the control pulse perspective, a rotation in the $xy$-plane of the Bloch sphere is implemented by modulating the relative phase $\phi$ of the $IQ$ envelope. We can rewrite the Hamiltonian from Equation~\ref{eq:hamiltonian-pulse-shape} as:
\begin{equation}
    H(t) = \frac{\gamma(t)}{2} \left[ \cos(\phi)\sigma_x + \sin(\phi)\sigma_y \right].
\end{equation}
The phase $\phi$ determines the rotation axis in the $xy$-plane. For a constant amplitude pulse $\gamma_c$ over a duration $T$, the unitary operator becomes:
\begin{equation}
    U = \exp\left(-i \frac{\gamma_c T}{2} \left[ \cos(\phi)\sigma_x + \sin(\phi)\sigma_y \right] \right).
\end{equation}
Thus, by changing only the phase $\phi$ of the control signal, one can rotate around any axis in the $xy$-plane without modifying the actual waveform. For instance, $\phi = 0$ corresponds to a rotation around the $x$ axis, while $\phi = \pi/2$ corresponds to a rotation around the $y$ axis.

Since $\phi$ sets the rotation axis, a $Z$ rotation can be implemented by adjusting the phase. A rotation with an arbitrary phase $\phi_0$ can be expressed in terms of a standard $R_x(\theta)$ rotation and virtual $Z$ gates:
\begin{equation}
    e^{-i \frac{\theta}{2} \left[ \cos(\phi_0)\sigma_x + \sin(\phi_0)\sigma_y \right]}
    = R_z(-\phi_0) R_x(\theta) R_z(\phi_0),
\end{equation}
which shows that a phase shift between pulses implements a virtual $Z$ rotation \cite{McKay2017}. This operation is performed entirely in software, without requiring any physical pulses, and is therefore considered noiseless and instantaneous.

We have described the theory and the calibration procedures necessary to implement single-qubit gates. Before addressing the more complex case of two-qubit gates, we first introduce a key preliminary step: determining the transition frequency of each qubit. This frequency calibration is a prerequisite for gate operations, as it determines the carrier frequency of control pulses.

\subsection{Finding the Qubit Frequency}\label{subsec:qubit-frequency}

Superconducting qubits, such as transmons, are physical systems whose energy levels resemble those of a quantum harmonic oscillator--an idealized system with equally spaced energy levels. However, transmons differ from the ideal case in a crucial way: the spacing between their energy levels is not exactly uniform. This deviation from perfect regularity is known as anharmonicity, and in the case of transmons, it is small, so we say they are weakly anharmonic.

Despite having multiple energy levels, the dynamics of a transmon are typically restricted to the two lowest ones, $\ket{0}$ and $\ket{1}$. The energy gap between these levels defines the qubit transition frequency $\omega$, which determines the resonance condition for the microwave pulses used to implement single-qubit gates.

A commonly used model to describe this behavior is the Duffing oscillator, which captures the anharmonic nature of the energy level spacings. Its Hamiltonian in the rotating frame is given by~\cite[Section II.A]{Krantz2019}:
\begin{equation}\label{eq:duff}
    H = \omega a^\dagger a + \frac{\alpha}{2} a^\dagger a^\dagger a a,
\end{equation}
where $a$ ($a^\dagger$) are operators that lower (raise) the energy level of the system, $\omega$ is the transition frequency between $\ket{0}$ and $\ket{1}$, and $\alpha < 0$ is the anharmonicity. The negative sign ensures that transitions to higher levels like $\ket{2}$ are off-resonant, allowing the system to behave effectively as a qubit.

Precisely characterizing the frequency $\omega$ of each qubit is a crucial calibration step, as this frequency defines the carrier of the microwave pulses used to perform single-qubit rotations. Deviations from resonance result in phase accumulation errors and reduced gate fidelity.

Here we describe how to experimentally determine this frequency using spectroscopy techniques. The basic idea is to prepare the qubit in its ground state $\ket{0}$ and apply a low-amplitude microwave pulse of fixed duration at varying frequencies. After each pulse, a measurement is performed to determine the probability of the qubit being excited to the $\ket{1}$ state.

When the drive frequency is far detuned from the qubit transition frequency $\omega$, the applied pulse does not effectively couple to the qubit, and the system remains in the ground state. This is because off-resonant drives fail to match the energy gap required for the $\ket{0} \rightarrow \ket{1}$ transition, resulting in negligible population transfer. As the drive frequency approaches resonance, the probability of excitation increases, reaching a maximum when the frequency precisely matches the energy difference $\hbar \omega$.

\begin{figure}[!h]
    \centering
    \includegraphics[width=0.9\linewidth]{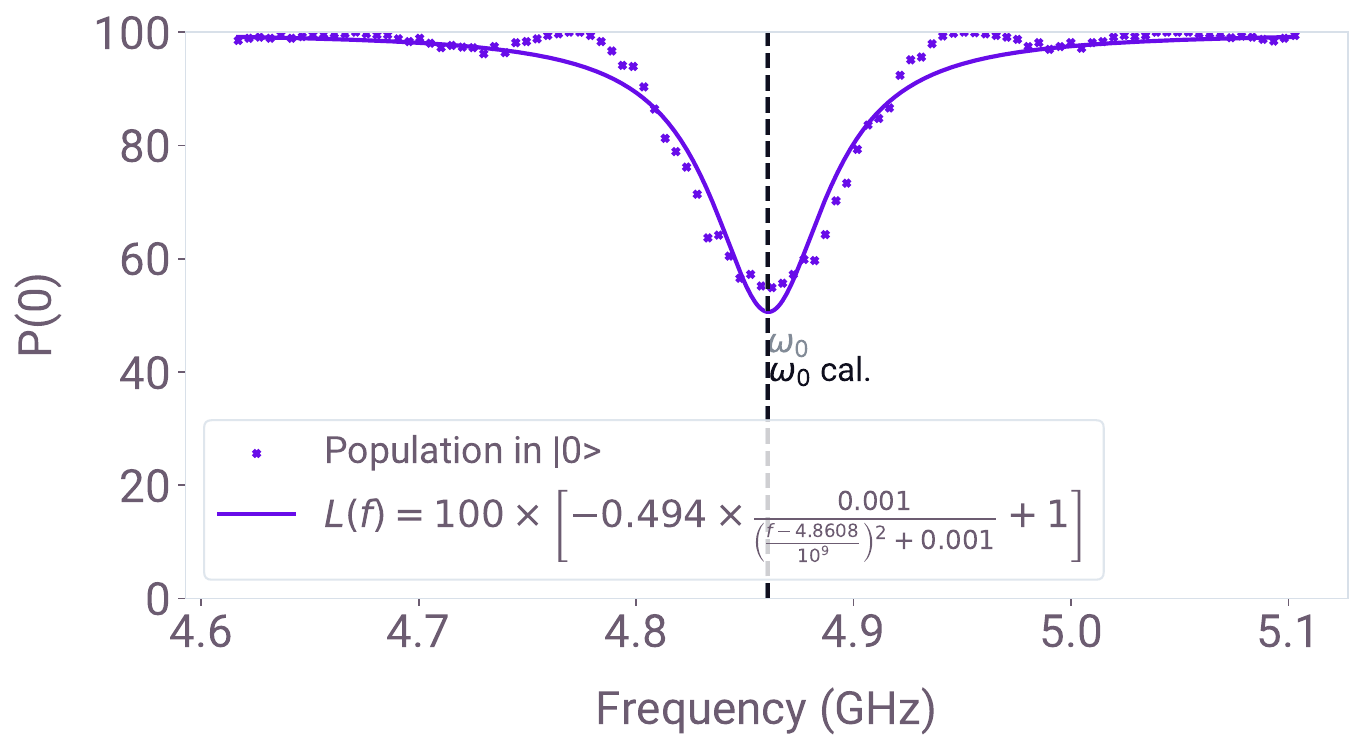}
    \caption{Frequency spectroscopy of a superconducting qubit obtained with the Qiskit Pulse simulator. The probability of measuring the ground state $|0\rangle$ is plotted as a function of the drive frequency. The dip corresponds to the qubit resonance at $\omega_0$ (gray dashed line), while $\omega_{0}^{\text{cal}}$ (black dashed line) indicates the calibrated frequency. The solid line is a Lorentzian fit.}
    \label{fig:frequency-sweep}
\end{figure}

This behavior results in a frequency-dependent excitation profile, as illustrated in Figure~\ref{fig:frequency-sweep}, with a characteristic Lorentzian shape centered around the true qubit frequency. The measured excitation probability as a function of drive frequency $f$ is typically fitted to a function of the form:
\begin{equation}
    \mathcal{L}(f) = -A \cdot \frac{\Gamma^2}{(f - \omega)^2 + \Gamma^2} + B,
\end{equation}
where $A$ is the amplitude of the curve, $B$ is an offset, $\omega$ is the resonance frequency, and $\Gamma$ is the linewidth (related to the qubit's decoherence and energy relaxation times). The center of this Lorentzian peak provides an accurate estimate of the qubit transition frequency $\omega$, which is then used to calibrate the carrier frequency of control pulses.

Beyond enabling single-qubit control, knowledge of each qubit's transition frequency also plays a central role in multi-qubit gate calibration. In particular, the relative detuning between qubits determines the strength and nature of their effective coupling, whether mediated by fixed-frequency interactions--as in the cross-resonance gate--or by tunable elements, as in $\mathrm{iSWAP}$ and $\mathrm{CPHASE}$ gates. In the next section, we show how multi-qubit gates can be implemented.

\subsection{Multi-Qubit Gates Calibration}\label{subsec:multi-qubit-gates-calibration}

Unlike single-qubit gates--which are typically implemented using direct microwave drives--there are multiple techniques available for realizing multi-qubit gates in superconducting qubits. The choice of technique depends mostly on the underlying hardware design, such as whether the qubits are fixed-frequency or frequency-tunable.

Up to this point, we have abstracted many low-level hardware details to focus on the software and control aspects of gate calibration. However, as we move to multi-qubit interactions, these hardware characteristics become more relevant and must be explicitly considered. In particular, the ability to tune the qubit frequency or exploit specific types of couplings determines which two-qubit gate protocols can be implemented.

Some qubits are equipped with a flux line, which allows dynamic tuning of this frequency by adjusting the magnetic flux threading the qubit loop. This tunability enables gates that rely on frequency detuning, such as the $\mathrm{iSWAP}$ (Section~\ref{subsubsec:iswap}) and $\mathrm{CPHASE}$ (Section~\ref{subsubsec:cphase}) gates.

On the other hand, qubits without flux lines--\textit{i.e.}, fixed-frequency qubits--typically implement multi-qubit gates using the cross-resonance technique (Section~\ref{subsubsec:cross-resonance}). In this method, a microwave pulse resonant with one qubit is applied through the control line of another qubit, inducing a controllable interaction between them.

Each of these methods introduces its own calibration challenges, such as mitigating flux noise in tunable qubits or dealing with spurious interactions in cross-resonance schemes. In this section, we explore how these multi-qubit gates are implemented and calibrated in practice.

\subsubsection{iSWAP}\label{subsubsec:iswap}

The $\mathrm{iSWAP}$ quantum gate~\cite{iSWAP} arises naturally from the coherent interaction between two coupled qubits. The dominant interaction term between two capacitively coupled qubits can be expressed as follows~\cite[Section IV.E.1]{Krantz2019}:
\begin{equation}\label{eq:iswap-int-hamiltonian}
    H = g \, \sigma_y^{(1)} \otimes \sigma_y^{(2)},
\end{equation}
where $g$ is the coupling strength, and $\sigma_y^{(i)}$ denotes the Pauli-$Y$ operator acting on qubit $i$.

To better understand the dynamics induced by this interaction, it is useful to express the Hamiltonian in terms of the ladder operators $\sigma_\pm = (\sigma_x \pm i\sigma_y)/2$. In the interaction picture and applying the rotating wave approximation (RWA), fast-oscillating terms are neglected under the assumption that they average out over time due to their high frequency. This leads to the simplified interaction Hamiltonian:
\begin{equation}
    H = g \left( e^{i\delta\omega_{12}t} \, \sigma_+^{(1)} \sigma_-^{(2)} + e^{-i\delta\omega_{12}t} \, \sigma_-^{(1)} \sigma_+^{(2)} \right),
\end{equation}
where $\delta\omega_{12} = \omega_1 - \omega_2$ is the frequency detuning between the two qubits. This expression describes an energy-conserving exchange of excitations between the two qubits.

If we tune the qubits such that their frequencies match, \textit{i.e.}, $\omega_1 = \omega_2$, then $\delta\omega_{12} = 0$ and the Hamiltonian becomes time-independent:
\begin{equation}
    H = g \left( \sigma_+^{(1)} \sigma_-^{(2)} + \sigma_-^{(1)} \sigma_+^{(2)} \right) = \frac{g}{2} \left( \sigma_x^{(1)} \sigma_x^{(2)} + \sigma_y^{(1)} \sigma_y^{(2)} \right).
\end{equation}
This is the well-known XY-type interaction Hamiltonian, which preserves the total excitation number and is responsible for coherent population exchange between the qubits.

The time evolution under this Hamiltonian is governed by the unitary operator:
\begin{equation}
    U(t) = \exp\left( -i \frac{g}{2} \left( \sigma_x \otimes \sigma_x + \sigma_y \otimes \sigma_y \right)t \right) =
    \begin{bmatrix}
        1 & 0          & 0          & 0 \\
        0 & \cos(gt)   & -i\sin(gt) & 0 \\
        0 & -i\sin(gt) & \cos(gt)   & 0 \\
        0 & 0          & 0          & 1 \\
    \end{bmatrix}.
\end{equation}

By allowing this interaction to evolve for a time $t' = \pi/2g$, the unitary becomes:
\begin{equation}
    U(\pi/2g) =
    \begin{bmatrix}
        1 & 0  & 0  & 0 \\
        0 & 0  & -i & 0 \\
        0 & -i & 0  & 0 \\
        0 & 0  & 0  & 1 \\
    \end{bmatrix}
    \equiv \mathrm{iSWAP}.
\end{equation}

Alternatively, evolving for half this time $t'' = \pi/4g$ leads to the so-called square-root of iSWAP gate:
\begin{equation}
    U(\pi/4g) =
    \begin{bmatrix}
        1 & 0                   & 0                   & 0 \\
        0 & \frac{1}{\sqrt{2}}  & -\frac{i}{\sqrt{2}} & 0 \\
        0 & -\frac{i}{\sqrt{2}} & \frac{1}{\sqrt{2}}  & 0 \\
        0 & 0                   & 0                   & 1 \\
    \end{bmatrix}
    \equiv \sqrt{\mathrm{iSWAP}}.
\end{equation}

The $\mathrm{iSWAP}$ gate performs the following transformations on the two-qubit computational basis:
\begin{align*}
    \ket{00} & \rightarrow \ket{00},   \\
    \ket{01} & \rightarrow -i\ket{10}, \\
    \ket{10} & \rightarrow -i\ket{01}, \\
    \ket{11} & \rightarrow \ket{11}.
\end{align*}

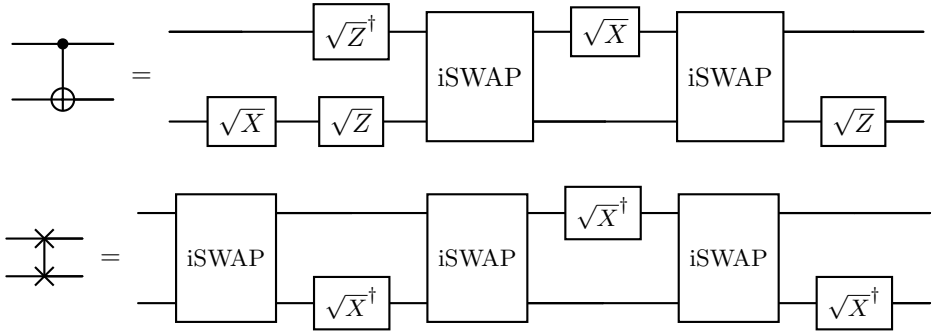
\begin{figure}[ht]
    \centering

    \begin{quantikz}
        & \ctrl{1} &\qw \\
        & \targ{}  &\qw
    \end{quantikz}
    =
    \begin{quantikz}
        &  \qw               & \gate{\sqrt{Z}^\dagger}  & \gate[2]{\mathrm{iSWAP}}
        & \gate{\sqrt{X}} & \gate[2]{\mathrm{iSWAP}} &       \qw          &\qw
        \\
        & \gate{\sqrt{X}} & \gate{\sqrt{Z}}          &
        &         \qw        &                          & \gate{\sqrt{Z}} &\qw
    \end{quantikz}

    \vspace{0.8em}

    \begin{quantikz}
        & \swap{1}  & \qw\\
        & \swap{-1} &\qw
    \end{quantikz}
    =
    \small{%
        \begin{quantikz}
            & \gate[2]{\mathrm{iSWAP}} &    \qw                      & \gate[2]{\mathrm{iSWAP}}
            & \gate{\sqrt{X}^\dagger}  & \gate[2]{\mathrm{iSWAP}} &                   \qw      &\qw
            \\
            &                          & \gate{\sqrt{X}^\dagger}  &
            &                    \qw      &                          & \gate{\sqrt{X}^\dagger} &\qw
        \end{quantikz}}

    \caption{$\mathrm{CNOT}$ and $\mathrm{SWAP}$ decomposition using $\mathrm{iSWAP}$ gates and single-qubit rotations~\cite{iSWAP}.}
    \label{fig:iswap-decompositions}
\end{figure}

This gate effectively swaps the $\ket{01}$ and $\ket{10}$ states up to a phase factor of $-i$, while leaving $\ket{00}$ and $\ket{11}$ unchanged. Because of this behavior, the iSWAP gate is often used as an entangling gate in superconducting qubit platforms. Furthermore, gates such as $\mathrm{CNOT}$ and $\mathrm{SWAP}$ can be constructed from $\mathrm{iSWAP}$ and single-qubit rotations, as illustrated in Figure~\ref{fig:iswap-decompositions}.

\subsubsection{CPHASE}\label{subsubsec:cphase}

To implement a controlled-phase ($\mathrm{CPHASE}$) interaction, we use a technique conceptually similar to that used for the $\mathrm{iSWAP}$ gate, but now explicitly take into account the presence of higher energy levels. Recall that for flux-tunable qubits, the qubit frequency can be adjusted by applying an external magnetic flux, thereby changing the energy level spacings. Up to this point, however, we have considered only single-qubit effects and neglected qubit-qubit interaction.

In reality, when two coupled quantum states are brought near resonance, their energy levels repel each other due to hybridization. This phenomenon is known as an avoided level crossing. For example, when we increase the flux applied to qubit 1 to bring its frequency close to that of qubit 2 (\textit{i.e.}, $\omega_1 \approx \omega_2$), the energy of $\ket{10}$ decreases until it approaches that of $\ket{01}$. However, due to the coupling, the two states do not cross but instead split, indicating a strong resonant exchange interaction. At this operating point, coherent oscillations between $\ket{10}$ and $\ket{01}$ can be used to implement the $\mathrm{iSWAP}$ gate. Moreover, because of the avoided crossing, further increasing the flux shifts the frequency of qubit 2 rather than that of qubit 1.

To implement a $\mathrm{CPHASE}$ gate, we exploit a different avoided crossing, this time between the $\ket{11}$ and $\ket{20}$ states. Starting from the computational state $\ket{11}$, we slowly -- that is, adiabatically -- vary the flux applied to qubit 1 toward the avoided crossing point at $\phi_{\mathrm{CPHASE}}$, where the states $\ket{11}$ and $\ket{20}$ hybridize. After waiting for a duration $\tau$ at this point, we return adiabatically to the initial flux bias $\phi_0 = 0$. The final state remains $\ket{11}$ but acquires an additional phase relative to the other computational basis states. This conditional phase enables the implementation of the $\mathrm{CPHASE}$ gate.

This behavior can be understood using the adiabatic theorem~\cite[Section 5.6.2]{sakurai2020}. According to the theorem, if a system is initially prepared in an eigenstate of a time-dependent Hamiltonian $H(t)$ and the Hamiltonian varies sufficiently slowly, the system will remain in the corresponding instantaneous eigenstate throughout the evolution, acquiring only a phase factor. In the computational basis, and assuming that each basis state remains approximately in its instantaneous eigenstate throughout the trajectory $l(\tau)$, this results in the following unitary operator:
\begin{equation}
    U_{\text{ad}} =
    \begin{bmatrix}
        1 & 0                   & 0                   & 0                   \\
        0 & e^{i\theta_{01}(l)} & 0                   & 0                   \\
        0 & 0                   & e^{i\theta_{10}(l)} & 0                   \\
        0 & 0                   & 0                   & e^{i\theta_{11}(l)}
    \end{bmatrix},
\end{equation}
where the phase acquired by each state is given by:
\begin{equation}
    \theta_{ij}(l(\tau)) = \int_0^\tau \omega_{ij}[l(t)] \,dt,
\end{equation}
with $\omega_{ij}$ denoting the instantaneous energy of state $\ket{ij}$ during the flux trajectory $l(t)$ of qubit 1.

Now define $\zeta = \omega_{11} - (\omega_{01} + \omega_{10})$ as the phase accumulation rate difference between the $\ket{11}$ state and the combined $\ket{01} + \ket{10}$ states. This detuning arises due to the repulsion of the $\ket{11}$ level caused by the nearby $\ket{20}$ level at the avoided crossing.

By choosing a flux trajectory $l_\pi$ such that the integrated detuning equals $\pi$,
\[
    \int_0^\tau \zeta[l_\pi(t)]\,dt = \theta_{11}(l_\pi) - \theta_{01}(l_\pi) - \theta_{10}(l_\pi) = \pi,
\]
the resulting unitary evolution becomes:
\begin{equation}
    U_{\text{ad}}
    =
    \begin{bmatrix}
        1 & 0                       & 0                       & 0                                                    \\
        0 & e^{i\theta_{01}(l_\pi)} & 0                       & 0                                                    \\
        0 & 0                       & e^{i\theta_{10}(l_\pi)} & 0                                                    \\
        0 & 0                       & 0                       & e^{i[\pi + \theta_{01}(l_\pi) + \theta_{10}(l_\pi)]}
    \end{bmatrix}.
\end{equation}

We can eliminate the accumulated single-qubit phases by applying appropriate virtual-$Z$ gates such that $\theta_{01}(l_\pi) = \theta_{10}(l_\pi) = 0$. This yields the final unitary:
\begin{equation}
    U_{\text{ad}} =
    \begin{bmatrix}
        1 & 0 & 0 & 0        \\
        0 & 1 & 0 & 0        \\
        0 & 0 & 1 & 0        \\
        0 & 0 & 0 & e^{i\pi}
    \end{bmatrix}
    =
    \begin{bmatrix}
        1 & 0 & 0 & 0  \\
        0 & 1 & 0 & 0  \\
        0 & 0 & 1 & 0  \\
        0 & 0 & 0 & -1
    \end{bmatrix}
    = \mathrm{CZ}.
\end{equation}

More generally, by engineering the flux pulse shape and duration, one can calibrate a wide range of controlled-phase gates, not limited to $\mathrm{CZ}$. In the next section, we explore a purely microwave-based two-qubit gate mechanism -- Cross-Resonance -- used primarily in fixed-frequency transmon architectures.

\subsubsection{Cross-Resonance}\label{subsubsec:cross-resonance}

When flux-tunable qubits are not available, an alternative method for implementing two-qubit entangling gates is the Cross-Resonance (CR) technique~\cite{CR}. The CR gate is a microwave-only scheme commonly used in fixed-frequency transmon architectures. The basic idea is to apply a microwave pulse on the control qubit at the transition frequency of the target qubit. Due to the qubit-qubit coupling, this induces a Rabi oscillation in the target qubit that depends on the state of the control qubit.

\begin{figure}[h]
    \centering
    \includegraphics[width=0.8\linewidth]{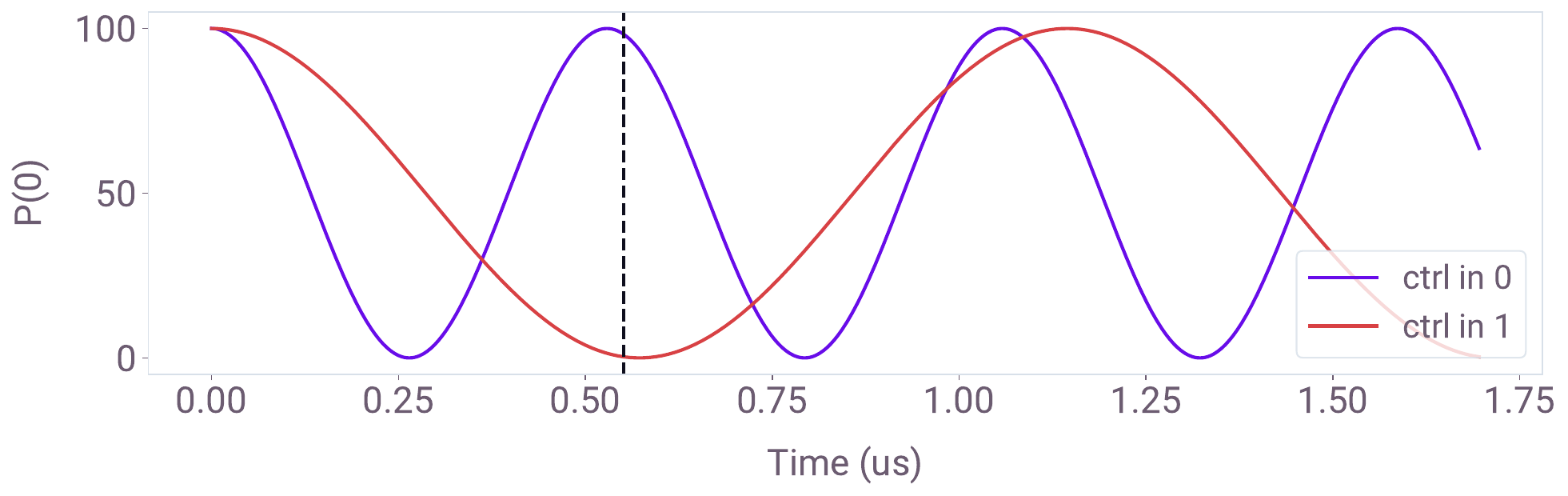}
    \caption{Rabi oscillations of the target qubit induced by a cross-resonance pulse applied to the control qubit, obtained with the Qiskit Pulse simulator. The oscillation frequency depends on whether the control qubit is in $\ket{0}$ (purple) or $\ket{1}$ (red). The point at which the oscillations reach opposite extrema (here, around \SI{0.55}{\mu\second}) is suitable for implementing a $\mathrm{CNOT}$ gate.}
    \label{fig:cross-resonance}
\end{figure}

Figure~\ref{fig:cross-resonance} shows the expectation value of $\braket{\sigma_z}$ for the target qubit as a function of time, conditioned on the control qubit being in state $\ket{0}$ (blue) or $\ket{1}$ (red). Notice that the Rabi oscillation frequencies differ depending on the control state. To implement a $\mathrm{CNOT}$ gate, we seek a moment in time where the target qubit remains in $\ket{0}$ when the control is in $\ket{0}$, and flips to $\ket{1}$ when the control is in $\ket{1}$--\textit{i.e.}, when the expectation values are +1 and –1, respectively. In this example, such synchronization occurs at $t = \SI{180}{\nano\second}$.

However, in practice, these Rabi frequencies do not always align so neatly. In many cases, several oscillations are required before a point of synchronization suitable for a $\mathrm{CNOT}$ occurs, which may increase the gate time and reduce fidelity.

\begin{figure}[h]
    \centering
    \includegraphics[width=0.8\linewidth]{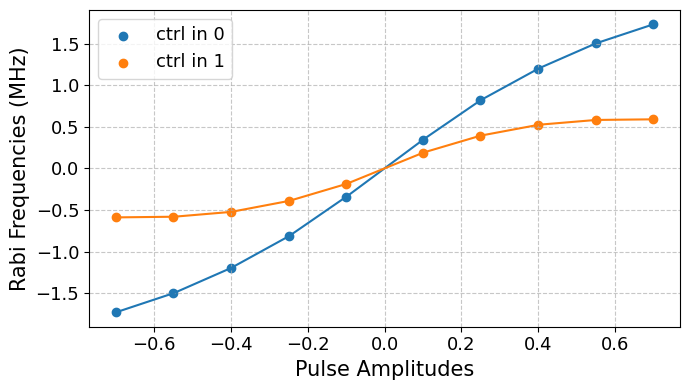}
    \caption{Example of a cross-resonance Rabi calibration obtained with the Qiskit Pulse simulator. The target qubit undergoes Rabi oscillations induced by a cross-resonance pulse while the control qubit is prepared in $\ket{0}$ (blue) and $\ket{1}$ (orange). These curves are used to estimate the drive amplitude and gate time for implementing a $\mathrm{CNOT}$.}
    \label{fig:cr-rabi-exp}
\end{figure}

To calibrate a cross-resonance gate, one should perform a Rabi experiment on the target qubit with the control qubit initialized in both $\ket{0}$ and $\ket{1}$, as described in Section~\ref{subsubsec:rabi-experiment}. The result should be similar to that shown in Figure~\ref{fig:cr-rabi-exp}, where the oscillation curves allow estimation of the Rabi frequencies conditioned on the control qubit state and pulse amplitude. From this, an appropriate drive amplitude can be chosen to minimize the synchronization time and optimize the $\mathrm{CR}$ gate duration. Note that this amplitude is typically fine-tuned beyond the initial estimate for optimal performance.

To understand the functioning of the cross-resonance (CR) gate, we begin by modeling the system as two coupled transmon qubits, each truncated to its two lowest energy levels, \(\ket{0}\) and \(\ket{1}\). In this reduced basis, the effective two-qubit Hamiltonian takes the form:
\begin{equation}
    H_{qq} = \sum_{i=1,2} \frac{\omega_i}{2} \sigma_z^{(i)} + g\, \sigma_y^{(1)} \otimes \sigma_y^{(2)},
\end{equation}
where \(\omega_i\) is the transition frequency of qubit \(i\), and \(g\) represents the strength of the coherent coupling between them.

We now include a drive applied to qubit 1, but at the frequency of qubit 2, which is the essence of the cross-resonance mechanism. The drive Hamiltonian is written as:
\begin{equation}
    H_d(t) = V(t)\, \sigma_x^{(1)},
\end{equation}
where \(V(t) = V_I(t) + V_Q(t)\), and the quadrature components are given by:
\[
    V_I(t) = A_I(t)\cos(\omega_2 t + \phi), \quad V_Q(t) = A_Q(t)\sin(\omega_2 t + \phi),
\]
as introduced in Section~\ref{subsec:pulse-modeling}. The total Hamiltonian of the driven system is:
\begin{equation}
    H = H_{qq} + H_d(t).
\end{equation}

To simplify the analysis and highlight the effective interaction induced on qubit 2, we apply the Schrieffer-Wolff transformation~\cite{SchriefferWolff}. This perturbative method block-diagonalizes the Hamiltonian to first order, removing off-resonant couplings and yielding an effective interaction Hamiltonian of the form
\begin{equation}
    \tilde{H}_d \approx V(t)\, \sigma_x^{(1)} + V(t)\, \frac{g}{\Delta_{12}}\, \sigma_z^{(1)} \otimes \sigma_x^{(2)} \,,
\end{equation}
where \(\Delta_{12} = \omega_1 - \omega_2\).

The second term, proportional to \(\sigma_z^{(1)} \otimes \sigma_x^{(2)}\), corresponds to the effective interaction responsible for the cross-resonance mechanism~\cite{SchriefferWolff}. When the drive is resonant with qubit 1 (\(\omega_d = \omega_1\)), the first term dominates and directly drives qubit 1. When the drive is tuned to qubit 2 (\(\omega_d = \omega_2\)), the second term becomes resonant, activating a conditional drive on qubit 2 that depends on the state of qubit 1. This conditional interaction enables the implementation of two-qubit gates such as the $\mathrm{CNOT}$.

Consequently, the unitary of the $\mathrm{CR}$ gate is:
\begin{equation}
    \mathrm{CR}(\theta) = e^{-i\frac{\theta}{2} \sigma_z \otimes \sigma_x} =
    \begin{bmatrix}
        \cos(\theta/2)   & -i\sin(\theta/2) & 0               & 0               \\
        -i\sin(\theta/2) & \cos(\theta/2)   & 0               & 0               \\
        0                & 0                & \cos(\theta/2)  & i\sin(\theta/2) \\
        0                & 0                & i\sin(\theta/2) & \cos(\theta/2)
    \end{bmatrix}.
\end{equation}

This matrix reveals that simply calibrating the angle $\theta$ does not yield a standard $\mathrm{CNOT}$ gate. Additional single-qubit rotations are required to complete the transformation. In particular, when $\theta = -\pi/2$, a $\mathrm{CNOT}$ can be constructed using the $\mathrm{CR}$ gate together with $\sqrt{X}$ and $\sqrt{Z}$ gates:

\begin{figure}[ht]
    \centering
    \begin{quantikz}
        & \ctrl{1} & \qw \\
        & \targ{}  & \qw
    \end{quantikz}
    =
    \begin{quantikz}
        & \gate{\sqrt{Z}} & \gate{\mathrm{CR_{-\pi/2}}} &\qw \\
        & \gate{\sqrt{X}} & \ctrl{-1} & \qw
    \end{quantikz}
    \caption{Decomposition of a $\mathrm{CNOT}$ gate using the $\mathrm{CR}_{-\pi/2}$ gate. Here, the drive is applied to the top (control) qubit, inducing a conditional $\sigma_x$ on the bottom (target) qubit.}

    \label{fig:cr-decompositions}
\end{figure}
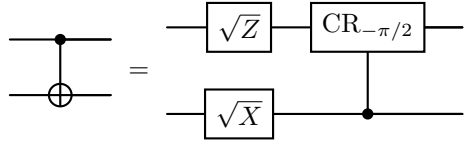

We have now explored three key two-qubit gate mechanisms--$\mathrm{iSWAP}$, $\mathrm{CPHASE}$, and Cross-Resonance--each suited to different qubit architectures and coupling strategies. With these entangling gates in place, we now turn to the final step of quantum processing: how measurement is performed and calibrated in superconducting qubit systems.

\subsection{Readout Calibration}\label{subsec:measurement-calibration}

To measure a qubit, we couple it to a readout resonator. Because of the dispersive interaction, the resonator frequency shifts depending on the qubit state. The basic idea of the dispersive readout is to apply a probe pulse to the resonator and analyze the reflected signal. Depending on the qubit state and the probe frequency, the amplitude and phase of the reflected signal are modified, allowing us to infer the qubit state.

In order to understand the behavior of the resonator coupled to the qubit~\cite{Koch2007}, consider the following Hamiltonian:
\begin{equation}
    H=\omega_r\left(a^\dagger a + \frac{1}{2}\right) + \frac{\omega_q}{2} \sigma_z + g\left(\sigma_+ a + \sigma_- a^\dagger\right),
\end{equation}
where $\omega_r$ is the resonator frequency, $\omega_q$ is the qubit frequency, and $g$ is the qubit–resonator coupling strength.

In the dispersive limit, where $\Delta = |\omega_q - \omega_r| \gg g$, the qubit and the resonator do not exchange energy resonantly, but instead cause small shifts in each other's frequencies. To analyze this regime, we apply the Schrieffer-Wolff transformation \cite{Koch2007}. This leads to the effective Hamiltonian:
\begin{equation}
    H_{\mathrm{eff}} = (\omega_r + \chi \sigma_z)\left(a^\dagger a + \frac{1}{2}\right) + \frac{\tilde{\omega}_q}{2} \sigma_z + \text{const.},
\end{equation}
where $\chi = \frac{g^2}{\Delta}$ is the dispersive shift, and $\tilde{\omega}_q = \omega_q + \chi$ is the Lamb-shifted qubit frequency.

This effective Hamiltonian shows that the resonator frequency becomes conditional on the qubit state. When the qubit is in $\ket{0}$ or $\ket{1}$, the resonator effectively oscillates at $\omega_r \pm \chi$, respectively. This state-dependent frequency shift is the key principle behind dispersive readout: by sending a probe signal near $\omega_r$ and observing its phase and amplitude response, one can infer the qubit state. At the same time, the qubit frequency is renormalized due to vacuum fluctuations of the resonator field, a correction known as the Lamb shift.

In practice, the measured quantity is the reflection coefficient $S_{11}$ of the probe tone. As the probe frequency is varied across the resonance, both the amplitude and phase of $S_{11}$ change in a way that depends on the qubit state. Typically, the reflected amplitude is reduced when the qubit is in one state and remains near unity in the other, while the corresponding phases differ significantly at specific detunings. The most common operating point is to fix the probe frequency at the bare resonator frequency $\omega_r$, where the two qubit states produce the largest phase separation in the reflected signal.

Another way to visualize the information is in the complex plane, where $S_{11}$ is represented as a vector with amplitude and phase. For each qubit state, repeated measurements form distinct clusters in this plane. The separation between these clusters quantifies the readout fidelity: the larger the separation, the easier it is to discriminate between $\ket{0}$ and $\ket{1}$.

Finally, to convert the analog readout signal into a digital qubit outcome, a classification algorithm is applied. A simple threshold on the measured quadrature components is often sufficient, but more sophisticated methods, such as maximum likelihood estimation or machine learning classifiers, can further improve readout accuracy.

\section{Final Remarks}\label{sec:conclusion}

In this work, we have detailed the architecture and components of a comprehensive, full-stack quantum software platform, from the high-level programming interface provided by Ket down to the physical pulse-level control of superconducting qubits. The development of such a complete stack is crucial for bridging the significant gap between abstract quantum algorithms and the complex, noisy physical hardware of the NISQ era. Mastering the entire compilation and execution flow provides a powerful tool for co-designing hardware and software, optimizing performance, and accelerating the path toward practical quantum applications.

This work represents a significant step forward for the quantum computing ecosystem. The platform's potential for technological impact was recently underscored when it was awarded first place in the prestigious SBC Innovation Seal 2025, an award from the Brazilian Computer Society (SBC) that recognizes academic research with a high potential to become impactful technology. This positions the platform as a cornerstone of a sovereign, national quantum infrastructure in Brazil, empowering the nation's scientific community and industry.

Nevertheless, this platform is a foundation upon which much can be built. We identify several key avenues for future expansion and improvement. The immediate priorities include the integration of more advanced error mitigation techniques, which are essential for extracting meaningful results from current hardware. Looking further ahead, incorporating the principles of quantum error correction will be necessary for the transition to fault-tolerant computing. The domain of quantum circuit optimization also offers vast room for enhancement, with opportunities to implement more sophisticated, hardware-aware optimization passes and explore advanced resynthesis techniques.

Looking ahead, our vision for the platform involves expanding the stack's capabilities at both the highest and lowest levels of abstraction. At the application level, we will develop domain-specific libraries to provide powerful tools for researchers in fields like quantum chemistry~\cite{Tilly2022} and finance~\cite{hermanQuantumComputingFinance2023}. Concurrently, we will push downwards towards the hardware interface by integrating pulse-level programming capabilities~\cite{alexanderQiskitPulseProgramming2020} and support for direct Arbitrary Waveform Generator (AWG) management~\cite{stefanazziQICKQuantumInstrumentation2022}. To foster a collaborative and transparent research environment, the core components of this stack are available at \url{https://quantumket.org} as an open-source platform, encouraging community contributions and accelerating the pace of innovation. This open model will not only benefit the research community but also serve as a valuable educational tool, helping to train the next generation of quantum scientists and engineers.

\backmatter

\bmhead{Acknowledgements}

ECRR acknowledges the Coordenação de Aperfeiçoamento de Pessoal de Nível Superior - CAPES, Finance Code 001;
EID, JM, and ECRR acknowledge the Conselho Nacional de Desenvolvimento Científico e Tecnológico - CNPq through grant number \mbox{409673/2022-6};
JM, EID, and ECRR acknowledge the Fundação de Amparo à Pesquisa e Inovação do Estado de Santa Catarina - FAPESC through Project FAPESC TR nº 2024TR002672;
EID acknowledges financial support from the National Institute for Science and Technology of Quantum Information (CNPq, INCT-IQ 465469/2014-0);
EWL acknowledges financial support from the Fundação de Amparo à Pesquisa e Inovação do Estado de Santa Catarina (FAPESC) under Public Call No. 24/2025.

\bibliography{sn-bibliography}

\end{document}